\documentclass{IEEEtran}
\usepackage{cite}
\usepackage{amsmath,amssymb,amsfonts}
\usepackage{algorithmic}
\usepackage{graphicx}
\usepackage{textcomp}

\usepackage{esint} 
\begin{document}
\title{Huygens' principle and field equivalence relations for cylindrical 
enclosing surfaces
}
\author{Andrey~Osipov,~\IEEEmembership{Senior~Member,~IEEE,} and 
Sergei~Tretyakov,~\IEEEmembership{Life~Fellow,~IEEE}%
\thanks{A.~V.~Osipov is with the Microwaves and Radar Institute, German Aerospace Center (DLR), 82234 Weßling, Germany (email: andrey.osipov@dlr.de).}%
\thanks{S.~A.~Tretyakov is with the Department of Electronics and Nanoengineering, School of Electrical Engineering, Aalto University, 02150 Espoo, Finland.}%
\thanks{This work was supported in part by the Research Council of Finland within the RCF-DoD Future Information Architecture for IoT initiative, grant no. 365679, Research Council of Finland grant no. 371367.}}

\maketitle

\begin{abstract}\label{abstract}
Frequency-domain Huygens' principle and equivalence relations for an infinitely long cylindrical
surface
enclosing the field sources or scatterers
are addressed.
Assuming 
the dependence   of the fields
along the cylinder axis $z$  as \(\exp\left( - jk_{z}z \right)\), 
where 
\(k_{z}\) is 
an arbitrary constant,
we derive three equivalent versions of 
line-integral representations
for the fields in free space at any point outside the 
enclosing cylinder. 
In one of the forms, 
similarly to the Helmholtz-Kirchhoff scalar diffraction
theory, the integrals contain a longitudinal field component, 
\(E_{z}\) or \(H_{z}\), and its
normal derivative. 
Another presented form contains longitudinal and normal field components, similarly to the three-dimensional  boundary integral representations by Stratton 
and
Chu. This form 
eliminates
the need to calculate the field derivatives. Furthermore, we present a two-dimensional version of 
the three-dimensional Schelkunoff-Franz representation, which contains only tangential components of the fields, complies with the field equivalence theorem
and can be regarded as a rigorous formulation of Huygens' principle for a cylindrical 
enclosing surface.
All three forms of the line-equivalence relation are exact and describe the fields through equivalent field distributions on a line enclosing 
the cross section of
the source region.  
A physical interpretation of Huygens' principle 
in terms of conical waves emanated by virtual linear sources on the cylindrical surface enclosing the true sources is given.
Specialization 
of the relations to the intermediate and far-field zones are presented. The derived expressions are applicable for studies of scattering and radiation in a very general class of structures that are infinite and periodic along one direction.

\end{abstract}


\section{Introduction}\label{Introduction}
Huygens' 
principle 
states that any external and induced field
sources located in a closed spatial region can be
replaced with
equivalent sources on 
any
surface
enclosing the original sources \cite[Ch. 1]{Huygens}.
The validity of this principle 
does not depend on 
the media filling the space and the sources outside 
the enclosing surface \cite{Lindell}. The equivalent sources are electric and magnetic surface currents that are defined by the   
tangential components of the electric and magnetic fields on the enclosing surface.
The fundamental reason for these properties is the mathematical form of the
Maxwell-Faraday and Amp\`{e}r-Maxwell laws that
relate the time variation of the electric field to the 
curl of the magnetic field and vice versa.

As soon as the Huygens surface currents are found, the fields outside the enclosing surface can be calculated using conventional techniques. In this paper, we 
assume free space outside of Huygens' surface.
In this case, 
Huygens' principle can also be formulated as the field equivalence theorem 
\cite[Sec. 6.14]{Schelkunoff},
\cite[Sec. 1.7]{Collin}, 
\cite[Sec. 3.5]{Harrington},
\cite{Sch-1951},
\cite[Sec. 7.8]{Balanis}.
In contrast to the concept of Huygens' sources, these formulations assume a homogeneous medium
outside the Huygens surface. The derivations 
are based on the uniqueness theorem for electromagnetic fields and state that the field of actual sources can be replaced by fictitious (or virtual) sources distributed over the surface enclosing the actual sources. The 
virtual sources are related to the tangential components of the electric and magnetic fields on the boundary, i.e. to Huygens' sources.

Practical applications of the equivalence theorem require a corresponding representation of fields at any point through their boundary values.
By using Green's functions the fields in a volume free of sources can be expressed
as surface integrals of their boundary values.
For scalar wave fields, the corresponding integral relation is known as
Helmholtz-Kirchhoff integral theorem (see \cite[Ch. 2]{BaCo}, \cite[Sec. 8.13]{Stratton} and \cite[Sec. 8.3.1]{BoWo}). 
For electromagnetic fields
in the frequency domain, 
two versions of the boundary integral relations are
available: the Stratton-Chu formula \cite[Sec. 8.14]{Stratton}, 
\cite{StCh-1939}
and the Schelkunoff-Franz 
formula
\cite[Sec. 6.1]{Schelkunoff}, \cite{Sch-1936}, \cite{Sch-1939}, 
\cite{Franz-1948} 
and \cite[Sec. 25]{Franz}\footnote{The authors of \cite{StCh-1939} and \cite{Franz-1948} seemed to be unaware of the earlier work \cite{Sch-1936}. Stratton and Chu acknowledged this fact a month later (see their note at the end of \cite{Sch-1939}) and emphasized the relation to Schelkunoff's formula. Franz used a different approach -- dyadic Green's functions rather than electromagnetic potentials -- to obtain a formula identical to that from \cite{Sch-1936}. 
}.
The latter contains only the tangential field components and therefore can be seen as one based on true Huygens' sources.

There is an extensive literature on 3D equivalence principle and related integral formulas, e.g. 
\cite[§ 16]{Tai},
\cite[Sec. 5.3]{Kong}, 
\cite{Chen-1989},
\cite[§ 1.4]{Chew},  
\cite[Sec. 1.6]{PRM}, 
\cite{ReRS-2000}
and \cite{Boo-2003}.
The principle finds applications in many practical engineering problems, including the design of antennas, radomes and metasurfaces, e.g. 
\cite{Nar-2022,Sug-2024,Wu-2024,Bar-2026}.

In a number of applications, it is convenient to assume that the
structure consists of cylindrical bodies infinite in one
direction, e.g. a line current, a wire,  
a transmission line, an open waveguide,
a periodic array, 
a section of an aircraft, 
etc., thus making the problem two-dimensional. Recently, two-dimensional models have been extensively used in studies of scanning reflectarrays and reconfigurable intelligent services, e.g. \cite{Popov,Elefth,Li,Sravan25}.
The 3D boundary integral relations assume, however, a compact enclosing surface and do
not apply to the case of infinitely long structures. A special treatment
of the problems with cylindrical geometry is therefore necessary.

A line-integral representation for a scalar solution of the
two-dimensional Helmholtz equation
has been derived in  the past by using a two-dimensional version of
the divergence theorem (also known as Gauss', Green's or
Ostrogradsky's theorem) \cite[Sec. 6.2]{BaCo}, \cite{Weber}.
A closed integration line separates the sources from the observation point,
and the integral
gives the solution of the Helmholtz equation 
in the source-free region
and vanishes in the complementary region. This result can be seen as an equivalence
relation for the electromagnetic fields independent of the \(z\)
coordinate.

Generalization to the electromagnetic case, in which the 
fields depend on \(z\), can
be carried out by considering the dependence in the form of the factor
\(\exp\left( - jk_{z}z \right)\), where \(k_{z}\) is a constant. For
example, \(k_{z} = k\cos\beta\) for a plane wave incident under the
angle \(\beta\) to the \(z\) axis. 
For a waveguide, $k_z=\mp j \gamma$, where $\gamma$ is the propagation constant.
The fields with a general dependence on \(z\) can be expanded in a Fourier spectrum with \(k_{z}\) being
the spectral variable, 
e.g. in a series of Floquet modes when 
the cylindrical structures are periodic in  the $z$ direction.

Assuming 
the translational symmetry of the 
enclosing surface,
homogeneous and isotropic medium  
and the exponential dependence of the fields
in the outside, 
Maxwell's equations get split into two
polarization cases, TM and TE, each fully described by a single field
component, \(E_{z}\) or \(H_{z}\), respectively \cite[Sec. 6.1 and 6.2]{Stratton}. The components satisfy the
two-dimensional Helmholtz equations, which permits deriving a
line-equivalence 
relation
for every polarization
with the integration line being the directrix of the enclosing cylindrical surface. 
These
relations have been derived in \cite[Sec. 2.6.2]{OT}
by adjusting the approach from
\cite{Lindell} to the 2D Helmholtz equation.

The obtained 2D integral relations can be seen as 
a rigorous formulation of Huygens' principle
for an infinite cylindrical enclosing surface,
though such an interpretation
has not been put forward in the literature so far. A possible reason can
be that in contrast to the 3D electromagnetic equivalence theorem, which
can be expressed entirely in terms of the electric and magnetic field
components tangential to the integration surface, the 2D version derived in \cite[Sec. 2.6.2]{OT}
involves the normal derivatives of the field components
as in the Helmholtz-Kirchhoff integral theorem.

A further problem with the normal derivative is that if the fields on the
enclosing contour are obtained from a numerical solution, then numerical
evaluation of the derivative seems to be necessary, which could lead to
losses in accuracy 
when using the equivalence formula.

In this paper, three equivalence relations for 
infinite cylindrical enclosing surfaces and
the electromagnetic fields with the
dependence \(\exp(- jk_{z}z)\) are presented. 
In Section~\ref{sec:ER-1} 
the 2D divergence
formula 
is applied to the scalar solutions of the 2D Helmholtz equation, which
recovers the result obtained in \cite[Sec. 2.6.2]{OT} by a slightly different approach.
In Section~\ref{sec:ER-2}, a modification of the line-equivalence formula is presented,
which does not include the normal derivatives of the fields on the integration
line. 
Section~\ref{sec:Stratton-Chu} demonstrates that the Stratton-Chu integral representation recovers the line-equivalence relation from Section~\ref{sec:ER-2} in the limit of an infinite cylindrical 
enclosing surface. 
In Section~\ref{sec:Franz}, the Schelkunoff-Franz integral representation is 
transformed to a line-equivalence relation that includes only the tangential field components. 

The formulas apply to cylindrical 
enclosing surfaces
with arbitrarily shaped cross sections and
can be used for extending the fields 
obtained by a numerical procedure
on the boundary of a computational domain into the exterior of the
domain, including the infinitely distant point. 
The line-integral representations can be significantly simplified when the observation point is moved at a sufficient distance from the integration contour. 
Section~\ref{sec:intermediate-region} presents simplified expressions for an observer
located at a distance greater than a wavelength from the integration line but not in the far-field region yet. 
The specification of the derived line-equivalence formulas to the 2D far-field limit is presented in Section~\ref{sec:far-field}. The presented formulas describe the far-field characteristics of cylindrical geometries (scattering amplitudes, scattering widths).
Section~\ref{sec:numerical-illustrations} gives numerical examples of the application of the presented line-equivalence relations to the field radiated by a line source and the field scattered by a homogeneous dielectric cylinder.

The time factor \(\exp(j\omega t)\) is supposed and omitted
throughout.

\section{Derivation of the equivalence relation in 2D}\label{sec:ER-1}

\begin{figure}
    \centering
    \includegraphics[width=0.9\linewidth]{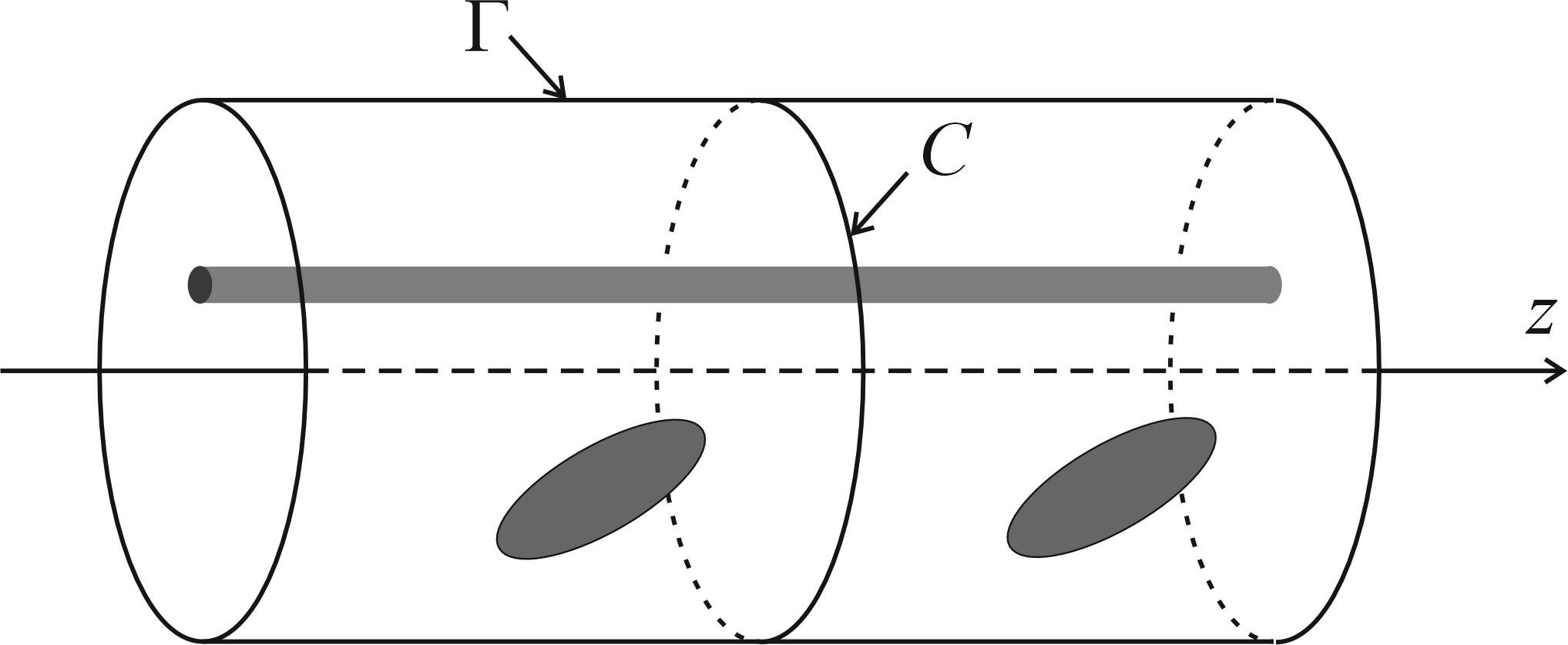}
        \caption{An infinite cylindrical surface $\Gamma$ along the $z$ axis enclosing variously shaped sources or/and scatterers 
    (here a wire and patches of an array); 
    the line $C$ is the directrix of $\Gamma$.}
    \label{fig:0}
\end{figure}

Consider a cylindrical 
surface $\Gamma$
of
infinite extent oriented along the \(z\) axis of a Cartesian coordinate system, 
which encloses a body or bodies of arbitrary material constitution and geometric shape, infinite cylindrical
or finite (Fig.~\ref{fig:0}).
The bodies can be either radiating (e.g. a line current) or scattering (e.g. a
metal or dielectric cylinder), in which case they are referred to as primary (or imprinted) and secondary sources, respectively. 
The medium outside the enclosing surface is assumed homogeneous with constant permittivity $\varepsilon$ and permeability $\mu$.
Because the fields can be expanded into Fourier series with respect to the $z$ coordinate, without loss of generality 
the dependence of the fields on \(z\) 
can be
assumed to be of the form \(\exp(- jk_{z}z)\), where \(k_{z}\) is a
constant.
Under these conditions, 
the fields outside of $\Gamma$ split up into two polarization cases, TM or E
case with \(H_{z} = 0\) and TE or H case with \(E_{z} = 0\),
and can be fully described in terms of a single component
\(E_{z}\) or \(H_{z}\), respectively (e.g. \cite[Sec. 6.1, 6.2]{Stratton}, 
\cite[Sec. 8]{Franz}).
The components
satisfy the two-dimensional Helmholtz equation 
\begin{equation}\label{eq:1}
\left( \nabla_{\bot}^{2} + k_{\bot}^{2} \right)U(\boldsymbol{\rho}) = 0 ,   
\end{equation}
where  
\(U(\boldsymbol{\rho}) = E_{z}\left( \boldsymbol{\rho},z \right)\exp(jk_{z}z)\)
(TM) or
\(U(\boldsymbol{\rho}) = H_{z}\left( \boldsymbol{\rho},z \right)\exp(jk_{z}z)\)
(TE), \(\boldsymbol{\rho}\) is the position-vector in the plane
perpendicular to the cylinder axis, \(\nabla_{\bot}^{2}\) is the
two-dimensional Laplacian, \(k_{\bot}^{2} = k^{2} - k_{z}^{2}\) and
\(k = \omega\sqrt{\varepsilon\mu}\) is the wave number. The rest
components of the fields can be determined from 
$E_z$ and $H_z$, 
e.g. \cite[Sec. 2.5.5]{OT}
for Cartesian and circular
cylindrical coordinates
or \eqref{eq:9a}--\eqref{eq:9d} for general orthogonal curvilinear coordinates.

The two-dimensional wave function
\(U\left( \boldsymbol{\rho} \right)\) is defined on an infinite plane 
in the exterior of 
the geometric cross section of the cylindrical 
region, in which the field sources are
located. 
For \(\rho \rightarrow \infty\),
\(U\left( \boldsymbol{\rho} \right)\) is required to be an outgoing wave,
which implies that
\(\left| U\left( \boldsymbol{\rho} \right) \right| \rightarrow 0\) if
Im \( k < 0\) (condition at infinity).

\begin{figure}
    \centering
    \includegraphics[width=0.9\linewidth]{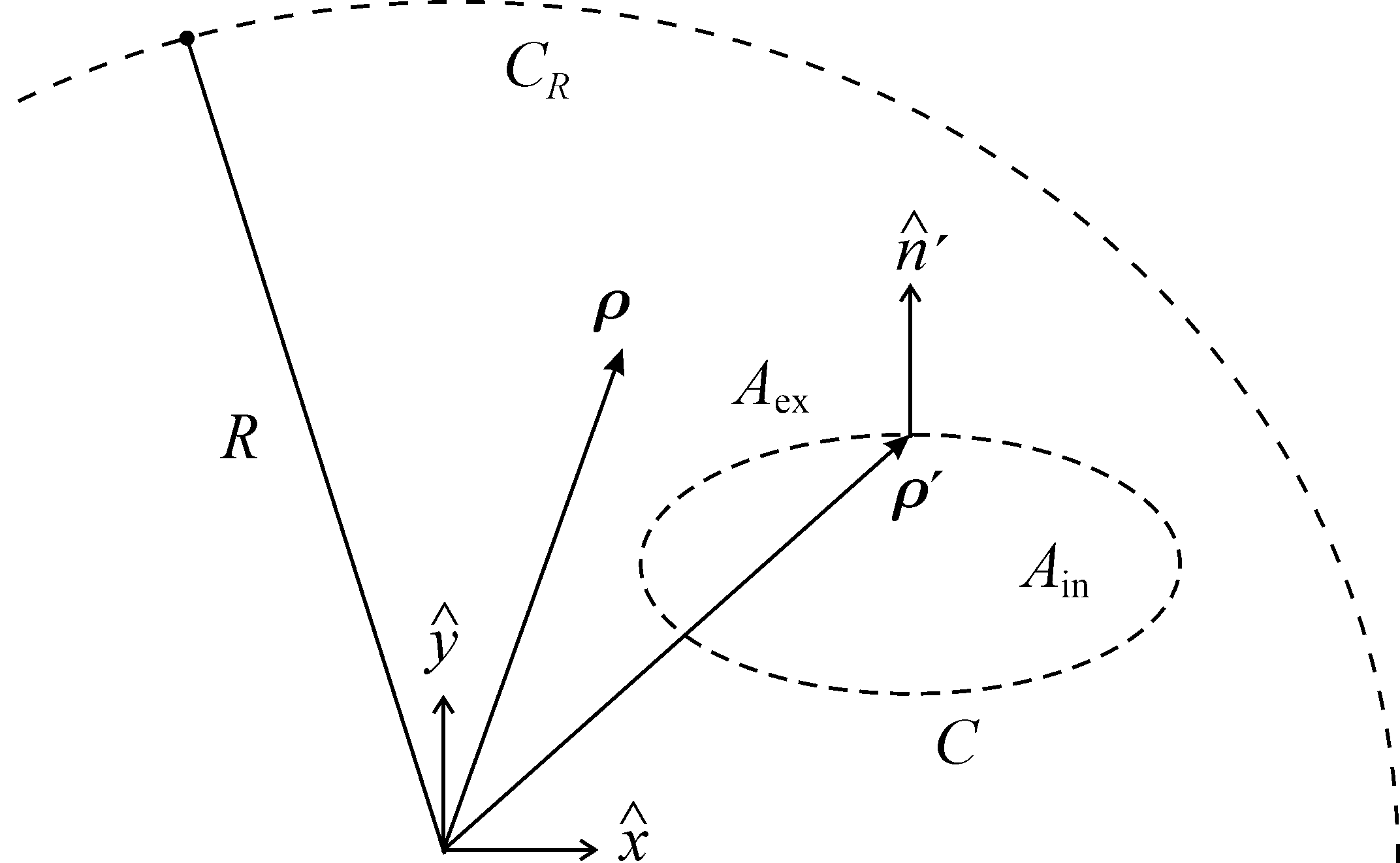}
        \caption{Geometry assumed in the formulation of the equivalence relation:
the region \(A_\mathrm{in}\) bounded by the contour \(C\) contains all the
sources; 
$A_\mathrm{ex}$ complements $A_\mathrm{in}$ to the full $x$-$y$ plane;
\(C_{R}\) is a circle of a large radius \(R\).}
    \label{fig:1}
\end{figure}

Denote the 
cross section of the cylindrical 
region containing all the considered sources by \(A_\mathrm{in}\), its boundary
by \(C\) and the region outside by \(A_\mathrm{ex}\) (Fig.~\ref{fig:1}). 
The line $C$ is therefore the directrix of the cylindrical region $\Gamma$ enclosing all sources.
In order to
express \(U\left( \boldsymbol{\rho} \right)\) in \(A_\mathrm{ex}\) through its
boundary values, the Green function of the infinite plane is required,
which is a solution of the equation
\begin{equation}\label{eq:2}
\left( \nabla_{\bot}^{2} + k_{\bot}^{2} \right)G_{2}\left( \boldsymbol{\rho},\boldsymbol{\rho}_{0},k_{\bot} \right) = - \delta\left( x - x_{0} \right)\delta\left( y - y_{0} \right)    
\end{equation}
with the delta source at a point
\(\boldsymbol{\rho}_{0}= x_{0}\hat{x} + y_{0}\hat{y}\). The
solution of \eqref{eq:2}, which is compliant with the condition at infinity, is given by
\begin{equation}\label{eq:3}
G_{2}\left( \boldsymbol{\rho},\boldsymbol{\rho}_{0},k_{\bot} \right) = \frac{1}{4j}H_{0}^{(2)}\left( k_{\bot}\left| \boldsymbol{\rho} - \boldsymbol{\rho}_{0} \right| \right),    
\end{equation}
where \(H_{0}^{(2)}(\cdots)\) is the Hankel function and
\(\left| \boldsymbol{\rho} - \boldsymbol{\rho}_{0} \right|\) is the distance
between the points.

Multiply \eqref{eq:1} with
\(G_{2}\left( \boldsymbol{\rho},\boldsymbol{\rho}_{0},k_{\bot} \right)\),
\eqref{eq:2} with \(U\left( \boldsymbol{\rho} \right)\) and subtract from each other
to obtain the relation
\begin{multline}\label{eq:4}
G_{2}\left( \boldsymbol{\rho},\boldsymbol{\rho}_{0},k_{\bot} \right)\nabla_{\bot}^{2}U\left( \boldsymbol{\rho} \right) - U\left( \boldsymbol{\rho} \right)\nabla_{\bot}^{2}G_{2}\left( \boldsymbol{\rho},\boldsymbol{\rho}_{0},k_{\bot} \right) \\    
=U\left( \boldsymbol{\rho} \right)\delta\left( x - x_{0} \right)\delta\left( y - y_{0} \right).
\end{multline}
The left-hand side is the divergence on the two-dimensional
vector field
\begin{equation}\label{eq:5}
\boldsymbol{F} = G_{2}\left( \boldsymbol{\rho},\boldsymbol{\rho}_{0},k_{\bot} \right)\nabla_{\bot}U\left( \boldsymbol{\rho} \right) - U\left( \boldsymbol{\rho} \right)\nabla_{\bot}G_{2}\left( \boldsymbol{\rho},\boldsymbol{\rho}_{0},k_{\bot} \right),    
\end{equation}
implying that one can integrate \eqref{eq:4} over \(A_\mathrm{ex}\) and apply the two-dimensional
version of the divergence theorem \cite[Sec.~1.4]{MoFe}
\begin{equation}\label{eq:6}
\iint\limits_{A_\mathrm{ex}}\left( \nabla_{\bot} \cdot \boldsymbol{F} \right)\,dA = - \oint\limits_{C}\left( \boldsymbol{F} \cdot \hat{n} \right)\,dl,    
\end{equation}
where \(\hat{n}\) is the unit normal to the boundary \(C\) pointing to 
\(A_\mathrm{ex}\) (Fig.~\ref{fig:1}), \(dA\) is the infinitesimal
surface element in the surface integral and \(dl\) is the infinitesimal arc length in the curvilinear integral. 

A mathematically correct way of applying \eqref{eq:6} to an
infinite region \(A_\mathrm{ex}\) consists in integrating 
first
over its 
large but bounded part \(A_{ex,R}\) 
\begin{multline}\label{eq:7}
\oint\limits_{C \cup C_{R}}
 \Big[ U\left( \boldsymbol{\rho} \right)\frac{\partial}{\partial n}G_{2}\left( \boldsymbol{\rho},\boldsymbol{\rho}_{0}, k_{\bot} \right) 
 - G_{2}\left( \boldsymbol{\rho},\boldsymbol{\rho}_{0}, k_{\bot} \right)\frac{\partial}{\partial n}U\left( \boldsymbol{\rho} \right) \Big]\, dl \\
 = 
\begin{cases}
U\left( \boldsymbol{\rho}_{0} \right), & \text{if $\boldsymbol{\rho}_{0} \in A_{\mathrm{ex},R}$} \\
0, & \text{if $\boldsymbol{\rho}_{0} \in A_\mathrm{in}$}
\end{cases}
,
\end{multline}
where $\partial /\partial n$
is the derivative in the direction
of the unit normal \(\hat{n}\) and \(C_{R}\) is a circle of a large
radius \(R\) so as to enclose the source region \(A_\mathrm{in}\). Then,
letting \(R\) go to infinity and accounting for the condition at
infinity transform \eqref{eq:7} into 
\begin{multline}\label{eq:8}
\oint\limits_{C }\Big[ U\left( \boldsymbol{\rho}' \right)\frac{\partial}{\partial n'}G_{2}\left( \boldsymbol{\rho},\boldsymbol{\rho}', k_{\bot} \right)
- G_{2}\left( \boldsymbol{\rho},\boldsymbol{\rho}', k_{\bot} \right)\frac{\partial}{\partial n'}U\left( \boldsymbol{\rho}' \right) \Big]\, dl' \\
= 
\begin{cases}
U\left( \boldsymbol{\rho} \right), & \text{if $\boldsymbol{\rho} \in A_\mathrm{ex}$ } \\
0, & \text{if $\boldsymbol{\rho} \in A_\mathrm{in}$}
\end{cases}
,
\end{multline}
where the observation point is renamed to \(\boldsymbol{\rho}\),
the integration point to \(\boldsymbol{\rho}'\), and the dash means the
dependence on the position of the integration point, e.g.,
\({\hat{n}}' = \hat{n}\left( \boldsymbol{\rho}' \right)\).
The normal derivative of Green's function evaluates to
\begin{equation}\label{eq:8a}
\frac{\partial}{\partial n'}G_{2}\left( \boldsymbol{\rho},\boldsymbol{\rho}',k_{\bot} \right) =
\frac{k_{\bot}}{4j}
H_{1}^{(2)}\left( k_{\bot}\left| \boldsymbol{\rho} - \boldsymbol{\rho}' \right| \right)\,
\hat{n}'\cdot \hat{s}_\bot (\boldsymbol{\rho},\boldsymbol{\rho}'),  
\end{equation}
where 
\begin{equation}\label{eq:s}
\hat{s}_\bot (\boldsymbol{\rho},\boldsymbol{\rho}')=\frac{\boldsymbol{\rho}-\boldsymbol{\rho}'}{|\boldsymbol{\rho}-\boldsymbol{\rho}'|}    
\end{equation}
is the unit vector in the plane perpendicular to the axis of the cylindrical region 
$\Gamma$.

Specifying \eqref{eq:8} with 
$\boldsymbol{\rho} \in A_\mathrm{ex}$ to the $z$-components of the fields leads to the relations:
\begin{multline}\label{eq:Ez}
E_z(\boldsymbol{\rho},z)
= \oint\limits_{C } \Big[ 
E_z\left( \boldsymbol{\rho}',z \right)
\frac{\partial}{\partial n'}G_{2}\left( \boldsymbol{\rho},\boldsymbol{\rho}', k_{\bot} \right)  \\    
-  G_{2}\left( \boldsymbol{\rho},\boldsymbol{\rho}', k_{\bot} \right)\frac{\partial}{\partial n'} E_z\left( \boldsymbol{\rho}',z \right) \Big] \, dl'
\end{multline}
\begin{multline}\label{eq:Hz}
H_z(\boldsymbol{\rho},z)
=  \oint\limits_{C }\Big[ H_z\left(\boldsymbol{\rho}',z \right)
\frac{\partial}{\partial n'}G_{2}\left( \boldsymbol{\rho},\boldsymbol{\rho}', k_{\bot} \right)    \\
- 
G_{2}\left( \boldsymbol{\rho},\boldsymbol{\rho}', k_{\bot} \right)\frac{\partial}{\partial n'} H_z\left( \boldsymbol{\rho}',z \right) \Big] \, dl'.
\end{multline}
Equations \eqref{eq:Ez} and \eqref{eq:Hz} hold for every value of $z$
and are in fact independent of $z$ as the factors $\exp(-jk_zz)$ 
in both sides of the equations cancel out.
Other field components result from Maxwell's equations (see 
\eqref{eq:9a}--\eqref{eq:9d} below).

Equations  \eqref{eq:Ez} and \eqref{eq:Hz} can be seen as  
field equivalence relations
in the case of infinite cylindrical 
enclosing surface.  
They state that
the \(z\) component of the field, either \(E_{z}\) or \(H_{z}\), at any
point outside of the source region can be exactly expressed through its
value and the value of its normal derivative on the boundary of the
source region
at the same value of $z$. 
The boundary values can be seen as the
equivalent sources 
that produce
the same field outside \(A_\mathrm{in}\) as the real sources. Note that the
shape and
 extent of the source region \(A_\mathrm{in}\) can be freely chosen as
long as \(A_\mathrm{in}\) contains all the sources and the material parameters in
\(A_\mathrm{ex}\) are position-independent. Furthermore, the 
boundary \(C\) must not necessarily be smooth and can include corner
points, e.g. when \(A_\mathrm{in}\) is a square region. At the corner
points the normal vector \(\hat{n}\) is undefined but the contribution
of any single point to the integral is zero as long as the integrand is
integrable at that point. Finally, 
the left-hand sides of \eqref{eq:Ez} and \eqref{eq:Hz}
are equal to zero
for an observer in $A_\mathrm{in}$.

A complimentary formulation, in which the sources are located in $A_\mathrm{ex}$, is also possible. It has the same form as \eqref{eq:Ez} and \eqref{eq:Hz}, the only difference being that the line integrals vanish for an observer in $A_\mathrm{ex}$ and are equal to the fields due to the sources (with the sign reversed) when $\boldsymbol{\rho} \in A_\mathrm{in}$.

\section{A modified form of the equivalence relation}\label{sec:ER-2}

The equivalence formula \eqref{eq:8} can be used for extending a solution from a
closed line, at which it is known, to any point beyond, including the
infinitely distant point ($\rho\rightarrow \infty$).
The extension
requires, however, the knowledge of \(U\) and $\partial U/\partial n$
on the boundary, which can be a
problem when the boundary field is available in a numeric form because
numerical evaluation of the derivative may result in the loss of
accuracy. A way around would be the use of a Green function vanishing
on \(C\), which would eliminate the term with the normal
derivative from \eqref{eq:8}. The problem is however that such Green's functions
are unavailable for arbitrarily shaped boundaries and when available,
e.g. for a circle or ellipse, the corresponding analytical expressions
are much more complicated compared with the Green function of a full plane
\eqref{eq:3}. Another special case is that in which a scattering problem is
considered for a 
cylindrical
scatterer with the Neumann boundary condition
\(\partial U/\partial n = 0\) or a boundary condition of the form
\(\partial U/\partial n + \delta U = 0\), where \(\delta\) is a
known parameter. Then the 
surface $\Gamma$
can be chosen to coincide with the physical boundary of the scatterer,
and the boundary conditions can be used to exclude the normal derivative from
the integrand.

In this Section it will be shown that in the most general case, i.e. for
arbitrarily shaped line \(C\), the terms
$\partial E_z/\partial n$
and
$\partial H_z/\partial n$
can be expressed through the field
components normal and tangential to \(C\), which will make the modified
2D equivalence relation 
suitable for integration of numerical data.

The problem can be solved by writing down Maxwell's equations 
in the outside of $\Gamma$
in a
suitable curvilinear coordinate system. Let \(u_{1}(x,y)\),
\(u_{2}(x,y)\) and \(u_{3} = z\) be a right-handed system of orthogonal
curvilinear coordinates, and let \({\hat{u}}_{i}\) be a unit vector
normal to the coordinate surface \(u_{i}\), oriented toward increasing
\(u_{i}\), and such that
\({\hat{u}}_{1} \times {\hat{u}}_{2} = {\hat{u}}_{3}\),
where \({\hat{u}}_{3} = \hat{z}\). For any cylindrical geometry,
the metric coefficients \(h_{1}\), \(h_{2}\) and \(h_{3}\) are such that
\(h_{3} = 1\) and \(h_{1}\) and \(h_{2}\) do not depend on \(z\). The
vector fields \(\boldsymbol{E}\) and \(\boldsymbol{H}\) are given by
\(\boldsymbol{E} = E_{1}{\hat{u}}_{1}\)+\(E_{2}{\hat{u}}_{2} + E_{z}\hat{z}\)
and \(\boldsymbol{H} = H_{1}{\hat{u}}_{1}\)+
\(H_{2}{\hat{u}}_{2} + H_{z}\hat{z}\) with the components
\(E_{i} = {\hat{u}}_{i}\cdot\boldsymbol{E}\) and
\(H_{i} = {\hat{u}}_{i}\cdot\boldsymbol{H}\).

Choose the coordinates \(u_{1}(x,y)\) and \(u_{2}(x,y)\) such that the
line \(u_{1}(x,y) = const\) coincides with the line \(C\). The
coordinate system can be obtained, e.g., by conformal mapping of a
circle onto the region \(A_\mathrm{in}\).
Now, using $\partial /\partial u_3= - jk_{z}$,
write
the curls in the chosen curvilinear coordinates as 
\cite[Sec. 1.16, 6.1 and 6.2]{Stratton}
\begin{multline}\label{eq:4a}
\nabla \times \boldsymbol{E}  =  \frac{1}{h_{2}}\left( \frac{\partial E_{z}}{\partial u_{2}} + jk_{z}h_{2}E_{2} \right){\hat{u}}_{1}  \\    
 - \frac{1}{h_{1}}\left( jk_{z}h_{1}E_{1} + \frac{\partial E_{z}}{\partial u_{1}} \right){\hat{u}}_{2} \\
 +  \frac{1}{h_{1}h_{2}}\left[ \frac{\partial}{\partial u_{1}}\left( h_{2}E_{2} \right) - \frac{\partial}{\partial u_{2}}\left( h_{1}E_{1} \right) \right]\hat{z}
\end{multline}
\begin{multline}\label{eq:4b}
\nabla \times \boldsymbol{H}  =  \frac{1}{h_{2}}\left( \frac{\partial H_{z}}{\partial u_{2}} + jk_{z}h_{2}H_{2} \right){\hat{u}}_{1}  \\    
-  \frac{1}{h_{1}}\left( jk_{z}h_{1}H_{1} + \frac{\partial H_{z}}{\partial u_{1}} \right){\hat{u}}_{2} \\
+  \frac{1}{h_{1}h_{2}}\left[\frac{\partial}{\partial u_{1}}\left( h_{2}H_{2} \right) - \frac{\partial}{\partial u_{2}}\left( h_{1}H_{1} \right) \right]\hat{z}
\end{multline}
and insert these into Maxwell's equations to obtain
\begin{eqnarray}
 j\omega\varepsilon E_{1}   &=& \frac{1}{h_{2}}\left( \frac{\partial H_{z}}{\partial u_{2}} + jk_{z}h_{2}H_{2} \right)
 \label{eq:5a} \\
j\omega\varepsilon E_{2}     &=& - \frac{1}{h_{1}}\left( jk_{z}h_{1}H_{1} + \frac{\partial H_{z}}{\partial u_{1}} \right)
\label{eq:5b} \\
 j\omega\varepsilon E_{z}     &=& 
\frac{1}{h_{1}h_{2}}\left\lbrack \frac{\partial}{\partial u_{1}}\left( h_{2}H_{2} \right) - \frac{\partial}{\partial u_{2}}\left( h_{1}H_{1} \right) \right\rbrack
\label{eq:5c} \\
- j\omega\mu H_{1}       &=& 
\frac{1}{h_{2}}\left( \frac{\partial E_{z}}{\partial u_{2}} + jk_{z}h_{2}E_{2} \right)
\label{eq:5d} \\
- j\omega\mu H_{2}        &=& 
- \frac{1}{h_{1}}\left( jk_{z}h_{1}E_{1} + \frac{\partial E_{z}}{\partial u_{1}} \right)
\label{eq:5e} \\
- j\omega\mu H_{z}         &=& 
\frac{1}{h_{1}h_{2}}\left\lbrack \frac{\partial}{\partial u_{1}}\left( h_{2}E_{2} \right) - \frac{\partial}{\partial u_{2}}\left( h_{1}E_{1} \right) \right\rbrack .
\label{eq:5f} 
\end{eqnarray}

Noting that
\begin{equation}\label{eq:HE}
\nabla \cdot \boldsymbol{E} = \frac{1}{h_{1}h_{2}}\left\lbrack \frac{\partial}{\partial u_{1}}\left( h_{2}E_{1} \right) + \frac{\partial}{\partial u_{2}}\left( h_{1}E_{2} \right) \right\rbrack - jk_{z}E_{z}    
\end{equation}
and using the Maxwell equation
$
\nabla \cdot (\varepsilon\boldsymbol{E)} = 0
$
with the
position-independent \(\varepsilon\),
equations \eqref{eq:5d} and \eqref{eq:5e} can be used in \eqref{eq:5c} to
obtain the Helmholtz equation for \(E_{z}\)
\begin{equation}\label{eq:HE1}
\frac{1}{h_{1}h_{2}}\left\lbrack \frac{\partial}{\partial u_{1}}\left( \frac{h_{2}}{h_{1}}\frac{\partial E_{z}}{\partial u_{1}} \right) + \frac{\partial}{\partial u_{2}}\left( \frac{h_{1}}{h_{2}}\frac{\partial E_{z}}{\partial u_{2}} \right) \right\rbrack + k_{\bot}^{2}E_{z} = 0. 
\end{equation}

The Helmholtz equation for
\(H_{z}\)
\begin{equation}\label{eq:HH}
\frac{1}{h_{1}h_{2}}\left\lbrack \frac{\partial}{\partial u_{1}}\left( \frac{h_{2}}{h_{1}}\frac{\partial H_{z}}{\partial u_{1}} \right) + \frac{\partial}{\partial u_{2}}\left( \frac{h_{1}}{h_{2}}\frac{\partial H_{z}}{\partial u_{2}} \right) \right\rbrack + k_{\bot}^{2}H_{z} = 0  
\end{equation}
results from \eqref{eq:5a}, \eqref{eq:5b} and \eqref{eq:5f} in a similar manner.

Consider now 
\eqref{eq:5a}, \eqref{eq:5b}, \eqref{eq:5d} and \eqref{eq:5e}. These are four equations
involving eight quantities: \(E_{1}\), \(E_{2}\), \(H_{1}\), \(H_{2}\),
$\partial E_z/\partial u_1$, $\partial E_z/\partial u_2$,
$\partial H_z/\partial u_1$ and $\partial H_z/\partial u_2$.
So, on the one hand, one can
use them to express the components  \(E_{1}\), \(E_{2}\), \(H_{1}\) and
\(H_{2}\) through the derivatives of \(E_{z}\) and \(H_{z}\) as 
\begin{eqnarray}
E_{1} & = & \frac{- j}{k_{\bot}^{2}}\left( \frac{k_{z}}{h_{1}}\frac{\partial E_{z}}{\partial u_{1}} + \frac{\omega\mu}{h_{2}}\frac{\partial H_{z}}{\partial u_{2}} \right)
\label{eq:9a}    \\
E_{2} & = & \frac{- j}{k_{\bot}^{2}}\left( \frac{k_{z}}{h_{2}}\frac{\partial E_{z}}{\partial u_{2}} - \frac{\omega\mu}{h_{1}}\frac{\partial H_{z}}{\partial u_{1}} \right)
\label{eq:9b}    \\
H_{1} & = & \frac{- j}{k_{\bot}^{2}}\left( \frac{k_{z}}{h_{1}}\frac{\partial H_{z}}{\partial u_{1}} - \frac{\omega\varepsilon}{h_{2}}\frac{\partial E_{z}}{\partial u_{2}} \right)
\label{eq:9c}    \\
H_{2}  & = & \frac{- j}{k_{\bot}^{2}}\left( \frac{k_{z}}{h_{2}}\frac{\partial H_{z}}{\partial u_{2}} + \frac{\omega\varepsilon}{h_{1}}\frac{\partial E_{z}}{\partial u_{1}} \right).
\label{eq:9d}    
\end{eqnarray}
Equations \eqref{eq:9a}--\eqref{eq:9d} permit determining the rest field components from the basic components $E_z$ and $H_z$.

On the other hand, 
\eqref{eq:5a}, \eqref{eq:5b}, \eqref{eq:5d} and \eqref{eq:5e}
can be used to express the derivatives
through the components
\begin{eqnarray}
\frac{1}{h_{1}}\frac{\partial E_{z}}{\partial u_{1}}  & = & 
j\left( \omega\mu H_{2} - k_{z}E_{1} \right)
 \label{eq:10a} \\   
\frac{1}{h_{1}}\frac{\partial H_{z}}{\partial u_{1}}  & = & 
 - j\left( \omega\varepsilon E_{2} + k_{z}H_{1} \right)
 \label{eq:10b} \\   
\frac{1}{h_{2}}\frac{\partial E_{z}}{\partial u_{2}}  & = & 
- j\left( \omega\mu H_{1} + k_{z}E_{2} \right)
 \label{eq:10c} \\ 
\frac{1}{h_{2}}\frac{\partial H_{z}}{\partial u_{2}}  & = & 
 j\left( \omega\varepsilon E_{1} - k_{z}H_{2} \right).
 \label{eq:10d}   
\end{eqnarray}
Equations \eqref{eq:10a}--\eqref{eq:10d} relate the normal and tangential derivatives of the \(z\)
components to the normal and tangential components of the fields on
the 
the surface $\Gamma$. 
Choose \({\hat{u}}_{1} = \hat{n}\) and
\({\hat{u}}_{2} = \hat{t}\), where \(\hat{n}\) is the unit
normal pointing to 
the exterior of $\Gamma$
and \(\hat{t}\) is the unit tangent
such that \(\hat{n} \times \hat{t} = \hat{z}\). The normal
derivative can now be written as
\begin{equation}\label{eq:11}
\frac{1}{h_{1}}\frac{\partial f}{\partial u_{1}} = \left( \hat{n},\nabla f \right) = \frac{\partial f}{\partial n},    
\end{equation}
and it follows from \eqref{eq:10a} and \eqref{eq:10b} that
\begin{eqnarray}
\frac{\partial E_{z}}{\partial n} & = & 
j\left( \omega\mu H_{t} - k_{z}E_{n} \right)
 \label{eq:12a}   \\
\frac{\partial H_{z}}{\partial n}  & = & 
- j\left( \omega\varepsilon E_{t} + k_{z}H_{n} \right).
 \label{eq:12b}    
\end{eqnarray}
Here, $E_n=\hat n \cdot \boldsymbol{E}$, $E_t=\hat t \cdot \boldsymbol{E}$, 
$H_n=\hat n \cdot \boldsymbol{H}$ and $H_t=\hat t \cdot \boldsymbol{H}$. 

Thus, using \eqref{eq:12a} and \eqref{eq:12b} in 
\eqref{eq:Ez} and \eqref{eq:Hz}
leads to the modified equivalence relations:
\begin{multline}\label{eq:13a} 
E_{z}\left( \boldsymbol{\rho}, z \right)   =  \oint\limits_{C}
\Big\{
E_{z}\left( \boldsymbol{\rho}',z \right)\frac{\partial}{\partial n'}G_{2}\left( \boldsymbol{\rho},\boldsymbol{\rho}', k_{\bot} \right)   \\
- 
j\big[\omega\mu H_{t'}\left( \boldsymbol{\rho}',z \right) - 
k_{z}E_{n'}\left( \boldsymbol{\rho}',z \right) \big]
G_{2}\left( \boldsymbol{\rho},\boldsymbol{\rho}', k_{\bot} \right) 
\Big \}\, dl'
\end{multline}
\begin{multline}\label{eq:13b} 
H_{z}\left( \boldsymbol{\rho}, z \right)  =  \oint\limits_{C}
\Big\{ 
H_{z}\left( \boldsymbol{\rho}',z \right)\frac{\partial}{\partial n'}G_{2}\left( \boldsymbol{\rho},\boldsymbol{\rho}', k_{\bot} \right)  \\    
+ 
j\big[ \omega\varepsilon E_{t'}\left( \boldsymbol{\rho}',z \right) + 
k_{z}H_{n'}\left( \boldsymbol{\rho}',z \right) \big] 
G_{2}\left( \boldsymbol{\rho},\boldsymbol{\rho}', k_{\bot} \right) 
\Big \}\, dl'
\end{multline}
for TM and TE polarization respectively. Equations \eqref{eq:13a} and \eqref{eq:13b}
assume that the observation point \(\boldsymbol{\rho}\) is in the region
\(A_\mathrm{ex}\). With \(\boldsymbol{\rho} \in A_\mathrm{in}\) the left-hand sides are
zeros. In general, the equivalence relations include both the tangential
and the normal components. In the limit \(k_{z} = 0\) (fields independent of \(z\)), the normal components \(E_{n}\) and \(H_{n}\)
disappear from \eqref{eq:13a} and \eqref{eq:13b} and only tangential components remain.

\section{Stratton-Chu's representation in the 2D limit}\label{sec:Stratton-Chu}
Another way of deriving the line-equivalence relations is to use the 3D boundary integral representations of electromagnetic fields. The procedure requires two steps: going to the limit of an infinite cylindrical integration region and integration with respect to $z$ in the surface integrals.
There are two fundamental integral representations -- the one by Stratton and Chu 
and another by 
Schelkunoff and Franz
(see Section \ref{Introduction}).
In this Section, the Stratton-Chu 
representation
is considered, and it will be shown that the formulation leads to the line-equivalence relations \eqref{eq:13a} and \eqref{eq:13b}.

Stratton-Chu's formulas 
describe the fields radiated by 
the sources 
enclosed by a surface $\Omega$ and
are given by the relations 
\begin{multline}\label{eq:ST-1}
\boldsymbol{E}(\boldsymbol{r}) =  
\oiint\limits_\Omega 
\big[ 
-j\omega\mu G_0(\boldsymbol{r},\boldsymbol{r}')\hat{n}'\times\boldsymbol{H}' 
 + 
(\hat{n}'\cdot\boldsymbol{E}')\nabla'G_0(\boldsymbol{r},\boldsymbol{r}')
 \\
+ 
(\hat{n}'\times\boldsymbol{E}')\times\nabla'G_0(\boldsymbol{r},\boldsymbol{r}') 
 \big] \, d\Omega'
\end{multline}
\begin{multline}\label{eq:ST-2}
\boldsymbol{H}(\boldsymbol{r}) = 
\oiint\limits_\Omega 
\big[ 
j\omega\varepsilon G_0(\boldsymbol{r},\boldsymbol{r}')\hat{n}'\times\boldsymbol{E}' 
+ (\hat{n}'\cdot\boldsymbol{H}')\nabla'G_0(\boldsymbol{r},\boldsymbol{r}')
\\    
+ 
(\hat{n}'\times\boldsymbol{H}')\times\nabla'G_0(\boldsymbol{r},\boldsymbol{r}') 
\big] \, d\Omega',
\end{multline}
where 
$\boldsymbol{r}$ and $\boldsymbol{r}'$ are 3D position vectors of an observation and an integration point, respectively,
$\nabla'$ is the 3D nabla operator acting on the primed coordinates and
\begin{equation}\label{eq:G0}
G_0(\boldsymbol{r},\boldsymbol{r}')=\frac{e^{-jk|\boldsymbol{r}-\boldsymbol{r}'|}}{4\pi |\boldsymbol{r}-\boldsymbol{r}'|}.
\end{equation}
The vector products of the fields with the unit normal describe the field components tangential to $\Omega$ and are referred to as equivalent surface currents,
\begin{equation}\label{eq:EqC}
\boldsymbol{K}_\mathrm{e}=\hat{n}\times\boldsymbol{H},~~~
\boldsymbol{K}_\mathrm{m}=\boldsymbol{E}\times\hat{n},
\end{equation}
electric and magnetic, respectively.

It is sufficient to consider \eqref{eq:ST-1} for the electric field. The corresponding result for the magnetic field will follow from the duality of Maxwell's equations by the substitutions: $\varepsilon\to\mu$, $\mu\to\varepsilon$, $\boldsymbol{E}\to \boldsymbol{H}$, $\boldsymbol{H}\to -\boldsymbol{E}$, $\boldsymbol{K}_\mathrm{e}\to \boldsymbol{K}_\mathrm{m}$ and
 $\boldsymbol{K}_\mathrm{m}\to -\boldsymbol{K}_\mathrm{e}$.

Equations \eqref{eq:ST-1} and \eqref{eq:ST-2} assume a compact integration 
surface. 
To extend them to the case of an infinite cylindrical 
surface $\Gamma$, consider $\Omega$ as a cylindrical region of a finite length $L$ and let $L$ go to infinity. 
In the limit, 
the contributions from the ends of the cylindrical region 
are required
 to vanish, implying that the dependence of the fields on $z$ 
 can be assumed to 
 be $\exp(-jk_zz)$ as 
 for an infinite cylindrical 
 boundary, which permits separated integration with respect to $z'$ 
 in \eqref{eq:ST-1}. 
 In the analytical calculation that follows, it is convenient to assume that $k_z$ is real-valued. This limitation is removed in the final result as the factor $\exp(-jk_zz)$ is factored out.

 In the limit of the infinite cylindrical structure, 
 by using the relations 
\begin{equation}\label{eq:exp}
\boldsymbol{E}(\boldsymbol{\rho}',z')=e^{-jk_zz'}\boldsymbol{E}(\boldsymbol{\rho}',0),~
\boldsymbol{H}(\boldsymbol{\rho}',z')=e^{-jk_zz'}\boldsymbol{H}(\boldsymbol{\rho}',0)
\end{equation}
to extract the $z$-dependence from the boundary fields,
 \eqref{eq:ST-1} 
 can be represented as
\begin{multline}\label{eq:ST-3}
\boldsymbol{E}(\boldsymbol{\rho},z) =
-j\omega\mu\oint\limits_C I_0(\boldsymbol{r},\boldsymbol{\rho}')\, 
\boldsymbol{K}_\mathrm{e}(\boldsymbol{\rho}',0)\,dl' \\    
- \oint\limits_C 
\boldsymbol{K}_\mathrm{m}(\boldsymbol{\rho}',0)
\times
\boldsymbol{I}_1(\boldsymbol{r},\boldsymbol{\rho}')\,dl' \\
+ \oint\limits_C \boldsymbol{I}_1(\boldsymbol{r},\boldsymbol{\rho}')\, 
\hat{n}'\cdot \boldsymbol{E}(\boldsymbol{\rho}',0)
\,dl', 
\end{multline}
where
\begin{eqnarray}
I_0(\boldsymbol{r},\boldsymbol{\rho}')&=&\int\limits_{-\infty}^{+\infty} e^{-jk_zz'}
G_0(\boldsymbol{r},\boldsymbol{r}')\,dz' \label{eq:I0} \\
\boldsymbol{I}_1(\boldsymbol{r},\boldsymbol{\rho}') &=& \int\limits_{-\infty}^{+\infty} 
e^{-jk_zz'}\nabla'G_0(\boldsymbol{r},\boldsymbol{r}')\,dz'. \label{eq:I1}
\end{eqnarray}

Integral \eqref{eq:I0} can be evaluated analytically \cite[Eq. (2.412)]{OT} as
\begin{equation}\label{eq:I0-1}
I_0(\boldsymbol{r},\boldsymbol{\rho}') = e^{-jk_zz}G_2(\boldsymbol{\rho},\boldsymbol{\rho}',k_\bot). 
\end{equation}
Green's function \eqref{eq:G0} is symmetric with respect to its arguments, implying that
\begin{multline}\label{eq:I1-1}
\boldsymbol{I}_1(\boldsymbol{r},\boldsymbol{\rho}') 
= -\nabla I_0(\boldsymbol{r},\boldsymbol{\rho}')= \\
e^{-jk_zz}
\big[
jk_z  G_2(\boldsymbol{\rho},\boldsymbol{\rho}',k_\bot) \hat{z}
-\nabla_\bot G_2(\boldsymbol{\rho},\boldsymbol{\rho}',k_\bot)
\big].
\end{multline}

The line-integral representation for the magnetic field results from \eqref{eq:ST-3} by using duality,
\begin{multline}\label{eq:ST-4}
\boldsymbol{H}(\boldsymbol{\rho},z) =
-j\omega\varepsilon\oint\limits_C I_0(\boldsymbol{r},\boldsymbol{\rho}')\, 
\boldsymbol{K}_\mathrm{m}(\boldsymbol{\rho}',0)\,dl' \\    
+ \oint\limits_C 
\boldsymbol{K}_\mathrm{e}(\boldsymbol{\rho}',0)
\times
\boldsymbol{I}_1(\boldsymbol{r},\boldsymbol{\rho}')\,dl' \\
+ \oint\limits_C \boldsymbol{I}_1(\boldsymbol{r},\boldsymbol{\rho}')\, 
\hat{n}'\cdot \boldsymbol{H}(\boldsymbol{\rho}',0)
\,dl'.
\end{multline}
Equations \eqref{eq:ST-3} and \eqref{eq:ST-4} provide a version of the line-equivalence relations for cylindrical geometries.
From \eqref{eq:I0-1} and \eqref{eq:I1-1} it is seen that the factors $\exp(-jk_zz)$ in both sides of \eqref{eq:ST-3} and \eqref{eq:ST-4} can be canceled out,
which makes the derived formulas applicable to complex-valued $k_z$. On the other hand, the exponential factor in \eqref{eq:I0-1} and \eqref{eq:I1-1} implies that \eqref{eq:ST-3} and \eqref{eq:ST-4} relate the fields at the same value of $z$.

Consider the $z$ components of \eqref{eq:ST-3} and \eqref{eq:ST-4}.
Inserting \eqref{eq:I0-1} and \eqref{eq:I1-1} into \eqref{eq:ST-3} and \eqref{eq:ST-4}
and using the relations
\begin{equation}\label{eq:ncH}
\boldsymbol{K}_\mathrm{e}(\boldsymbol{\rho}',0)
=
-H_z(\boldsymbol{\rho}',0)\,\hat{t}' + H_{t'}(\boldsymbol{\rho}',0)\,\hat{z}     
\end{equation}
\begin{equation}\label{eq:ncE}    
\boldsymbol{K}_\mathrm{m}(\boldsymbol{\rho}',0)
=
E_z(\boldsymbol{\rho}',0)\,\hat{t}' - E_{t'}(\boldsymbol{\rho}',0)\,\hat{z} 
\end{equation}
\begin{equation}
\hat{z}\cdot\left[
\boldsymbol{K}_\mathrm{m}(\boldsymbol{\rho}',0)
\times\boldsymbol{I}_1(\boldsymbol{r},\boldsymbol{\rho}')
\right] 
=
-E_z(\boldsymbol{\rho}',0)\frac{\partial }{\partial n'}G_2(\boldsymbol{\rho},\boldsymbol{\rho}',k_\bot)    
\end{equation}
\begin{equation}
\hat{z}\cdot\left[
\boldsymbol{K}_\mathrm{e}(\boldsymbol{\rho}',0)
\times\boldsymbol{I}_1(\boldsymbol{r},\boldsymbol{\rho}')
\right]
=
H_z(\boldsymbol{\rho}',0)\frac{\partial }{\partial n'}G_2(\boldsymbol{\rho},\boldsymbol{\rho}',k_\bot)    
\end{equation}
recover expression \eqref{eq:13a} and \eqref{eq:13b} for the $z$ component of the electric and magnetic fields
if relations \eqref{eq:exp} are used.

\section{Schelkunoff-Franz' formula in the 2D limit}\label{sec:Franz}
The field equivalence theorem suggests the equivalent sources entirely in terms of  the field components tangential to the integration surface, namely $\hat{n}\times\boldsymbol{H}$ and $\hat{n}\times\boldsymbol{E}$. The Stratton-Chu formulas \eqref{eq:ST-1} and \eqref{eq:ST-2} include, however, the normal components of the fields $\hat{n}\cdot\boldsymbol{H}$ and $\hat{n}\cdot\boldsymbol{E}$, which means that despite of being exact, 
the formulas
are not based on the true Huygens' sources. The integral representations of fields published in 
\cite{Sch-1936}
and \cite{Franz-1948}
contain only the tangential components,
\begin{multline}\label{eq:F-1}
\boldsymbol{E}(\boldsymbol{r}) =  
\frac{1}{j\omega\varepsilon}
\left[k^2+\nabla (\nabla\cdot)\right]
\oiint\limits_\Omega  
G_0(\boldsymbol{r},\boldsymbol{r}')\,\hat{n}'\times\boldsymbol{H}' 
\,d\Omega'    \\
+ 
\nabla\times\oiint\limits_\Omega
G_0(\boldsymbol{r},\boldsymbol{r}')\, \hat{n}'\times\boldsymbol{E}'
 \, d\Omega'
\end{multline}
\begin{multline}\label{eq:F-2}
\boldsymbol{H}(\boldsymbol{r}) = 
-\frac{1}{j\omega\mu}
\left[k^2+\nabla (\nabla\cdot)\right]
\oiint\limits_\Omega  
G_0(\boldsymbol{r},\boldsymbol{r}')\,\hat{n}'\times\boldsymbol{E}' 
\,d\Omega'  \\
+ 
\nabla\times\oiint\limits_\Omega
G_0(\boldsymbol{r},\boldsymbol{r}')\, \hat{n}'\times\boldsymbol{H}'
 \, d\Omega'.
\end{multline}
Here, a compact surface $\Omega$ encloses all sources, and the observation point is located in the complementary source-free region. For an observation point inside $\Omega$ the left-hand sides of \eqref{eq:F-1} and \eqref{eq:F-2} are equal to zero.

The limit of \eqref{eq:F-1} and \eqref{eq:F-2} for infinite cylindrical regions can be studied along the same lines as in the case of Stratton-Chu representation (sec.~\ref{sec:Stratton-Chu}). Again, because of the duality, it is sufficient to consider any of the equations \eqref{eq:F-1} and \eqref{eq:F-2}. Assuming that $\Omega$ is a cylindrical region of a finite length $L$, letting $L$ go to infinity, requiring vanishing contributions from the ends of the region and using the formula \eqref{eq:I0-1} transform \eqref{eq:F-1}  into the following equation:
\begin{equation}\label{eq:F-3}
\boldsymbol{E}(\boldsymbol{\rho},z) =  
\frac{1}{j\omega\varepsilon}
\oint\limits_C  
\boldsymbol{A}(\boldsymbol{r},\boldsymbol{\rho}') 
\,dl'  - 
\oint\limits_C
\boldsymbol{B}(\boldsymbol{r},\boldsymbol{\rho}') 
 \, dl', 
\end{equation}
where
\begin{equation}\label{eq:F-A}
\boldsymbol{A}(\boldsymbol{r},\boldsymbol{\rho}')  = 
\left[k^2+\nabla (\nabla\cdot)\right]
\left[I_0(\boldsymbol{r},\boldsymbol{\rho}')\boldsymbol{K}_\mathrm{e}(\boldsymbol{\rho}',0)
\right]     
\end{equation}
\begin{equation}\label{eq:F-B}
\boldsymbol{B}(\boldsymbol{r},\boldsymbol{\rho}') = \nabla \times
\left[I_0(\boldsymbol{r},\boldsymbol{\rho}')\boldsymbol{K}_\mathrm{m}(\boldsymbol{\rho}',0)
\right]    
\end{equation}
and the definitions \eqref{eq:EqC} for the equivalent electric and magnetic currents are used.
In \eqref{eq:F-A} and \eqref{eq:F-B}, $\nabla$ operates on $\boldsymbol{\rho}$ and $z$ but not on $\boldsymbol{\rho}'$, implying that
\begin{equation}
\nabla\times \left[I_0(\boldsymbol{r},\boldsymbol{\rho}')\boldsymbol{K}_\mathrm{m}(\boldsymbol{\rho}',0)\right] = \nabla I_0(\boldsymbol{r},\boldsymbol{\rho}') \times \boldsymbol{K}_\mathrm{m}(\boldsymbol{\rho}',0)    
\end{equation}
\begin{equation}
\nabla (\nabla\cdot) \left[I_0(\boldsymbol{r},\boldsymbol{\rho}')\boldsymbol{K}_\mathrm{e}(\boldsymbol{\rho}',0)\right] =
\nabla \left[ \boldsymbol{K}_\mathrm{e}(\boldsymbol{\rho}',0) \cdot \nabla I_0(\boldsymbol{r},\boldsymbol{\rho}') \right],    
\end{equation}
and \eqref{eq:F-3} takes the form
\begin{multline}\label{eq:F-7}
\boldsymbol{E}(\boldsymbol{\rho},z) = 
\frac{1}{j\omega\varepsilon}
\oint\limits_C  
\Big\{ 
k^2 I_0(\boldsymbol{r},\boldsymbol{\rho}') \boldsymbol{K}_\mathrm{e}(\boldsymbol{\rho}',0)     \\
 -
\nabla \left[\boldsymbol{K}_\mathrm{e}(\boldsymbol{\rho}',0) \cdot \boldsymbol{I}_1(\boldsymbol{r},\boldsymbol{\rho}')\right]
\Big\}\, dl' \\
+ \oint\limits_C  
\boldsymbol{I}_1(\boldsymbol{r},\boldsymbol{\rho}') \times \boldsymbol{K}_\mathrm{m}(\boldsymbol{\rho}',0)\, dl'.
\end{multline}

The line-integral representation for $\boldsymbol{H}$ results from \eqref{eq:F-7}
by using the duality,
\begin{multline}\label{eq:F-8}
\boldsymbol{H}(\boldsymbol{\rho},z) = 
\frac{1}{j\omega\mu}
\oint\limits_C  
\Big\{ 
k^2 I_0(\boldsymbol{r},\boldsymbol{\rho}') \boldsymbol{K}_\mathrm{m}(\boldsymbol{\rho}',0)     \\
-
\nabla \left[\boldsymbol{K}_\mathrm{m}(\boldsymbol{\rho}',0) \cdot \boldsymbol{I}_1(\boldsymbol{r},\boldsymbol{\rho}')\right]
\Big\}\, dl' \\
- \oint\limits_C  
\boldsymbol{I}_1(\boldsymbol{r},\boldsymbol{\rho}') \times \boldsymbol{K}_\mathrm{e}(\boldsymbol{\rho}',0)\, dl'.
\end{multline}
In \eqref{eq:F-7} and \eqref{eq:F-8} the factor $\exp(-jk_zz)$ can be moved from $I_0$ and $\boldsymbol{I}_1$ to $K_\mathrm{e,m}(\boldsymbol{\rho}',0)$ to get a form in which the fields on both sides of the equations are taken at the same value of $z$.

Similarly to \eqref{eq:ST-3} and \eqref{eq:ST-4},
equations \eqref{eq:F-7} and \eqref{eq:F-8} are fully vectorial and describe all components of the fields. To compare with the line-equivalence relations
\eqref{eq:Ez}, \eqref{eq:Hz}, 
\eqref{eq:13a} and \eqref{eq:13b},
let us calculate the $z$ components of \eqref{eq:F-7} and \eqref{eq:F-8}.
The vector quantities \eqref{eq:F-A} and \eqref{eq:F-B} can be analytically calculated
by using \eqref{eq:3}, \eqref{eq:ncH} and \eqref{eq:ncE} with the following result for their $z$ components
\begin{multline}\label{eq:Az}
A_z(\boldsymbol{r},\boldsymbol{\rho}') = -e^{-jk_zz}\frac{k_\bot}{4}
\left[
jk_\bot H_{t'}(\boldsymbol{\rho}',0)H_0^{(2)}(k_\bot |\boldsymbol{\rho}-\boldsymbol{\rho}'|)
\right. \\ 
+\left. k_z H_z(\boldsymbol{\rho}',0)
H_1^{(2)}(k_\bot |\boldsymbol{\rho}-\boldsymbol{\rho}'|)\,
\hat{t}'\cdot \hat{s}_\bot (\boldsymbol{\rho},\boldsymbol{\rho}')
\right]
\end{multline}
\begin{multline}\label{eq:Bz}
B_z(\boldsymbol{r},\boldsymbol{\rho}') = \\ 
\displaystyle
-e^{-jk_zz}\frac{k_\bot}{4j}
E_z(\boldsymbol{\rho}',0)
H_1^{(2)}(k_\bot |\boldsymbol{\rho}-\boldsymbol{\rho}'|)\,
\hat{n}'\cdot \hat{s}_\bot (\boldsymbol{\rho},\boldsymbol{\rho}'),
\end{multline}
where $\hat{s}_\bot (\boldsymbol{\rho},\boldsymbol{\rho}')$ is defined in \eqref{eq:s}.

The $z$ projection of \eqref{eq:F-7} is therefore equal to
\begin{multline}\label{eq:F-4}
E_z(\boldsymbol{\rho},z) =  
-\frac{k_\bot}{4j}
\oint\limits_C 
\Big[
j\frac{k_\bot}{\omega\varepsilon} H_{t'}(\boldsymbol{\rho}',z)H_0^{(2)}(k_\bot |\boldsymbol{\rho}-\boldsymbol{\rho}'|) 
\\
+ 
\frac{k_z}{\omega\varepsilon} H_z(\boldsymbol{\rho}',z)
H_1^{(2)}(k_\bot |\boldsymbol{\rho}-\boldsymbol{\rho}'|)\,
\hat{t}'\cdot \hat{s}_\bot (\boldsymbol{\rho},\boldsymbol{\rho}')
\\
-
E_z(\boldsymbol{\rho}',z)
H_1^{(2)}(k_\bot |\boldsymbol{\rho}-\boldsymbol{\rho}'|)\,
\hat{n}'\cdot \hat{s}_\bot (\boldsymbol{\rho},\boldsymbol{\rho}')
\Big]\, dl'.
\end{multline}
Using \eqref{eq:8a} and 
\begin{equation}\label{eq:8b}
\frac{\partial}{\partial t'}G_{2}\left( \boldsymbol{\rho},\boldsymbol{\rho}',k_{\bot} \right) =
\frac{k_{\bot}}{4j}
H_{1}^{(2)}\left( k_{\bot}\left| \boldsymbol{\rho} - \boldsymbol{\rho}' \right| \right)\,
\hat{t}'\cdot \hat{s}_\bot (\boldsymbol{\rho},\boldsymbol{\rho}')    
\end{equation}
leads to the expressions:
\begin{multline}\label{eq:F-5} 
E_z(\boldsymbol{\rho},z) =  
\oint\limits_C 
\Big[
E_z(\boldsymbol{\rho}',z)\frac{\partial}{\partial n'}
G_{2}\left( \boldsymbol{\rho},\boldsymbol{\rho}',k_{\bot} \right) \\ 
- j\frac{k_\bot^2}{\omega\varepsilon} H_{t'}(\boldsymbol{\rho}',z)
G_{2}\left( \boldsymbol{\rho},\boldsymbol{\rho}',k_{\bot} \right) \\
- 
\frac{k_z}{\omega\varepsilon}
H_z(\boldsymbol{\rho}',z)
\frac{\partial}{\partial t'}G_{2}\left( \boldsymbol{\rho},\boldsymbol{\rho}',k_{\bot} \right)
\Big]\, dl'
\end{multline}
\begin{multline}\label{eq:F-6} 
H_z(\boldsymbol{\rho},z) =   
\oint\limits_C 
\Big[
H_z(\boldsymbol{\rho}',z)\frac{\partial}{\partial n'}
G_{2}\left( \boldsymbol{\rho},\boldsymbol{\rho}',k_{\bot} \right)
 \\
+ j\frac{k_\bot^2}{\omega\mu}
E_{t'}(\boldsymbol{\rho}',z)
G_{2}\left( \boldsymbol{\rho},\boldsymbol{\rho}',k_{\bot} \right) \\
+ 
\frac{k_z}{\omega\mu}
E_z(\boldsymbol{\rho}',z)
\frac{\partial}{\partial t'}G_{2}\left( \boldsymbol{\rho},\boldsymbol{\rho}',k_{\bot} \right)
\Big]\, dl'.
\end{multline}
It is seen that \eqref{eq:F-5} and \eqref{eq:F-6} are different from \eqref{eq:Ez}, \eqref{eq:Hz}, 
\eqref{eq:13a} and \eqref{eq:13b}
as they include only tangential field components. Moreover,  
\eqref{eq:F-5} and \eqref{eq:F-6} contain the tangential derivative of the Green function
in addition to the Green function and its normal derivative.

Representations \eqref{eq:F-7} and \eqref{eq:F-8} describe the fields outside of the source 
region
entirely in terms of the tangential field components
on the enclosing cylindrical surface $\Gamma$.
These relations can be seen as the line-equivalence relations that use truly Huygens' equivalent sources.
For an observation point inside the region $C$ the left-hand sides of 
\eqref{eq:F-7}, \eqref{eq:F-8},
\eqref{eq:F-5} and \eqref{eq:F-6} are equal to zero, which is the characteristic feature of Huygens' sources.

\section{The intermediate region}\label{sec:intermediate-region}
The line-equivalence relations derived before simplify as soon as the observation point $\boldsymbol{\rho}$ is removed from the integration contour $C$. The simplifications are associated with the following conditions:
\begin{equation}\label{eq:cond1}
d\gg \frac{1}{|k_\bot |}
\end{equation}
\begin{equation}\label{eq:cond2}
d\gg D
\end{equation}
\begin{equation}\label{eq:cond3}
d\gg |k_\bot | D^2,
\end{equation}
where $d =\min_{\boldsymbol{\rho}'\in C} |\boldsymbol{\rho}-\boldsymbol{\rho}'|$ is the shortest distance between the observer and the line $C$, and
\(D\) is the diameter of the region enclosed by $C$.

Condition \eqref{eq:cond1} is the most important one since it permits the most significant simplifications. For electrically small regions $A_\mathrm{in}$ ($k_\bot D\ll 1$), conditions 
\eqref{eq:cond2} and  \eqref{eq:cond3} are automatically met once \eqref{eq:cond1} is true.
For an electrically large region $A_\mathrm{in}$ ($k_\bot D\gg 1$), conditions 
\eqref{eq:cond2} and  \eqref{eq:cond3} are essential additional conditions that are satisfied at greater distances from $C$ than required by \eqref{eq:cond1}.
In the latter case, the approximations 
that follow
describe the fields in the intermediate region, i.e. at the distances between the distance at which \eqref{eq:cond1} is satisfied and the distance at which \eqref{eq:cond2}
and \eqref{eq:cond3} are met.

Note that \eqref{eq:cond1} and \eqref{eq:cond3} contain $k_\bot$ 
which
even in the high-frequency case ($k\gg 1$) can be small 
if
$k_z\approx \pm k$, i.e. for waves propagating along the axis of the cylindrical region.
On the other hand, if $k_z$ is small or zero, \eqref{eq:cond1} has the meaning of the high-frequency approximation ($d\gg \lambda/2\pi$), while condition \eqref{eq:cond3} is the far-zone condition for large scatterers or antennas ($d\gg D^2/\lambda$).

In this Section, we consider the situation, in which condition \eqref{eq:cond1} is satisfied, without any assumption about \eqref{eq:cond2} and \eqref{eq:cond3}. 
The Hankel function in the Green function \eqref{eq:3}
can be replaced by the large-argument approximation (e.g. \cite[Eq. (9.2.4)]{AbSt})
\begin{equation}\label{eq:G2-1}
G_{2}\left( \boldsymbol{\rho},\boldsymbol{\rho}',k_{\bot} \right)\approx   
G_2^\mathrm{as}\left( \boldsymbol{\rho},\boldsymbol{\rho}',k_{\bot} \right)=
\frac{e^{- jk_{\bot}|\boldsymbol{\rho}-\boldsymbol{\rho}'| - j\pi/4}}{\sqrt{8\pi k_{\bot}|\boldsymbol{\rho}-\boldsymbol{\rho}'|}}
\end{equation}
implying that
\begin{equation}\label{eq:G2-2}
I_0(\boldsymbol{r},\boldsymbol{\rho}')\approx e^{-jk_z z} G_2^\mathrm{as}\left( \boldsymbol{\rho},\boldsymbol{\rho}',k_{\bot} \right)
\end{equation}
and
\begin{equation}\label{eq:G2-3}
\boldsymbol{I}_1(\boldsymbol{r},\boldsymbol{\rho}')\approx 
j \boldsymbol{k}_\mathrm{s}'
e^{-jk_z z} G_2^\mathrm{as}\left( \boldsymbol{\rho},\boldsymbol{\rho}',k_{\bot} \right),
\end{equation}
where $\boldsymbol{k}_\mathrm{s}'= k_\bot \hat s_\bot(\boldsymbol{\rho},\boldsymbol{\rho}') + k_z \hat z$ is 
a
vector with the length $k$. 
Using 
\eqref{eq:exp} and 
\eqref{eq:G2-1}--\eqref{eq:G2-3} in the line-equivalence relations of the Stratton-Chu type, \eqref{eq:ST-3} and \eqref{eq:ST-4}, gives
\begin{multline}\label{eq:ST-5}
\boldsymbol{E}(\boldsymbol{\rho},z) \approx 
\frac{e^{ j \pi/4}}{\sqrt{8\pi k_{\bot}}}
\oint\limits_C 
\frac{e^{- jk_{\bot}|\boldsymbol{\rho}-\boldsymbol{\rho}'|}}{\sqrt{|\boldsymbol{\rho}-\boldsymbol{\rho}'|}}
\big[
-\omega\mu
\boldsymbol{K}_\mathrm{e}(\boldsymbol{\rho}',z)  \\
-\boldsymbol{K}_\mathrm{m}(\boldsymbol{\rho}',z)
\times
\boldsymbol{k}_\mathrm{s}'  + 
\hat{n}'\cdot \boldsymbol{E}(\boldsymbol{\rho}',z) \, \boldsymbol{k}_\mathrm{s}' 
\big]
\,dl'
\end{multline}
\begin{multline}\label{eq:ST-6}
\boldsymbol{H}(\boldsymbol{\rho},z) \approx 
\frac{e^{j\pi/4}}{\sqrt{8\pi k_{\bot}}}
\oint\limits_C 
\frac{e^{- jk_{\bot}|\boldsymbol{\rho}-\boldsymbol{\rho}'|}}{\sqrt{|\boldsymbol{\rho}-\boldsymbol{\rho}'|}}
\big[
-\omega\varepsilon
\boldsymbol{K}_\mathrm{m}(\boldsymbol{\rho}',z) \\ 
+\boldsymbol{K}_\mathrm{e}(\boldsymbol{\rho}',z)
\times
\boldsymbol{k}_\mathrm{s}'  + 
\hat{n}'\cdot \boldsymbol{H}(\boldsymbol{\rho}',z) \, \boldsymbol{k}_\mathrm{s}' 
\big]
\,dl'.
\end{multline}

The line-equivalence relations of the Schelkunoff-Franz type 
\eqref{eq:F-7} and \eqref{eq:F-8} simplify to 
\begin{multline}\label{eq:F-9}
\boldsymbol{E}(\boldsymbol{\rho},z) \approx 
\frac{e^{j\pi/4}}{\sqrt{8\pi k_{\bot}}}
\oint\limits_C 
\frac{e^{- jk_{\bot}|\boldsymbol{\rho}-\boldsymbol{\rho}'|}}{\sqrt{|\boldsymbol{\rho}-\boldsymbol{\rho}'|}}
\Big\{
\boldsymbol{k}_\mathrm{s}'\times\boldsymbol{K}_\mathrm{m}(\boldsymbol{\rho}',z)
\\    
+\frac{1}{\omega\varepsilon}
\boldsymbol{k}_\mathrm{s}'\times\left[\boldsymbol{k}_\mathrm{s}'\times\boldsymbol{K}_\mathrm{e}(\boldsymbol{\rho}',z)\right] 
\Big\}\,dl'
\end{multline}
\begin{multline}\label{eq:F-10}
\boldsymbol{H}(\boldsymbol{\rho},z) \approx 
\frac{e^{j\pi/4}}{\sqrt{8\pi k_{\bot}}}
\oint\limits_C 
\frac{e^{- jk_{\bot}|\boldsymbol{\rho}-\boldsymbol{\rho}'|}}{\sqrt{|\boldsymbol{\rho}-\boldsymbol{\rho}'|}}
\Big\{
-\boldsymbol{k}_\mathrm{s}'\times\boldsymbol{K}_\mathrm{e}(\boldsymbol{\rho}',z)
 \\
+\frac{1}{\omega\mu}
\boldsymbol{k}_\mathrm{s}'\times\left[\boldsymbol{k}_\mathrm{s}'\times\boldsymbol{K}_\mathrm{m}(\boldsymbol{\rho}',z)\right] 
\Big\}\,dl',
\end{multline}
where the relation 
\begin{equation}
\boldsymbol{k}_\mathrm{s}'\times\left[\boldsymbol{k}_\mathrm{s}'\times\boldsymbol{K}\right] =
\boldsymbol{k}_\mathrm{s}' \left(\boldsymbol{k}_\mathrm{s}'\cdot\boldsymbol{K}\right)-k^2 \boldsymbol{K}
\end{equation}
with $\boldsymbol{K}$ being either $\boldsymbol{K}_\mathrm{e}$ or $\boldsymbol{K}_\mathrm{m}$ has been used.

The approximate equations \eqref{eq:ST-5}--\eqref{eq:F-10} represent 
the fields of the enclosed sources as superpositions of conical waves originated at every point $\boldsymbol{\rho}'$ of the line $C$ with the amplitude factor as in a cylindrical wave and the phase function $k_\bot |\boldsymbol{\rho}-\boldsymbol{\rho}'|+k_zz $ that describes a conical wave front, e.g. \cite[Sec. 2.5]{OT}. So, Huygens' principle in the case of cylindrical geometries sees the outgoing wave as a superposition of conical waves emanated by every 
element of the enclosing cylindrical surface.

 \section{The far-field region}\label{sec:far-field}

The region, where all conditions \eqref{eq:cond1}--\eqref{eq:cond3} are met, will be referred to as the far-field region.
Let us take a look at the line-equivalence relations \eqref{eq:ST-5}--\eqref{eq:F-10} in the case that in addition to 
\eqref{eq:cond1}, further conditions are satisfied.
If \eqref{eq:cond2} is true, then
$|\boldsymbol{\rho}-\boldsymbol{\rho}'|$ in the denominator can be replaced with 
$|\boldsymbol{\rho}-\boldsymbol{\rho}_\mathrm{c}|$,
where $\boldsymbol{\rho}_\mathrm{c}$ is the geometric center of the region,
$\hat s_\bot(\boldsymbol{\rho},\boldsymbol{\rho}')$  with 
$\hat s_\bot^\mathrm{c} = \hat s_\bot(\boldsymbol{\rho},\boldsymbol{\rho}_\mathrm{c})$ 
and $\boldsymbol{k}_\mathrm{s}'$ with $\boldsymbol{k}_\mathrm{s}^\mathrm{c}= k_\bot \hat s_\bot^\mathrm{c} + k_z \hat z$, which do not depend on the position of the integration point.

At distances, at which the condition \eqref{eq:cond3} is satisfied, the approximation
\begin{equation}
|\boldsymbol{\rho}-\boldsymbol{\rho}'|=|\boldsymbol{\rho}-\boldsymbol{\rho}_\mathrm{c}|
- \hat s_\bot^\mathrm{c}\cdot (\boldsymbol{\rho}'-\boldsymbol{\rho}_\mathrm{c})+
O\left(D^2/d\right)
\end{equation}
can be used in the phase factors of \eqref{eq:ST-5}--\eqref{eq:F-10}, implying that
\begin{equation}
e^{- jk_{\bot}|\boldsymbol{\rho}-\boldsymbol{\rho}'|} \approx 
e^{- jk_{\bot}|\boldsymbol{\rho}-\boldsymbol{\rho}_\mathrm{c}|+jk_\bot
\hat s_\bot^\mathrm{c}\cdot (\boldsymbol{\rho}'-\boldsymbol{\rho}_\mathrm{c})}.  
\end{equation}
Note that for an electrically large region $A_\mathrm{in}$ ($|k_\bot| D\gg 1$), the condition \eqref{eq:cond3} is much more severe than \eqref{eq:cond1} and \eqref{eq:cond2}. 

With \eqref{eq:cond2} and \eqref{eq:cond3} satisfied, \eqref{eq:ST-5}--\eqref{eq:F-10} become
\begin{multline}\label{eq:ST-7}
\boldsymbol{E}(\boldsymbol{\rho},z) \approx 
\frac{e^{- jk_{\bot}|\boldsymbol{\rho}-\boldsymbol{\rho}_\mathrm{c}|+j\pi/4}}{\sqrt{8\pi k_{\bot}|\boldsymbol{\rho}-\boldsymbol{\rho}_\mathrm{c}|}}
\oint\limits_C 
e^{jk_\bot
\hat s_\bot^\mathrm{c}\cdot (\boldsymbol{\rho}'-\boldsymbol{\rho}_\mathrm{c})} \\
\bigg[
\boldsymbol{k}_\mathrm{s}^\mathrm{c}  \times \boldsymbol{K}_\mathrm{m}(\boldsymbol{\rho}',z)
-\omega\mu
\boldsymbol{K}_\mathrm{e}(\boldsymbol{\rho}',z) 
 + \, \hat{n}'\cdot \boldsymbol{E}(\boldsymbol{\rho}',z) \, \boldsymbol{k}_\mathrm{s}^\mathrm{c}  
\bigg] \,dl' 
\end{multline}
\begin{multline}\label{eq:ST-8}
\boldsymbol{H}(\boldsymbol{\rho},z) \approx 
\frac{e^{- jk_{\bot}|\boldsymbol{\rho}-\boldsymbol{\rho}_\mathrm{c}|+j\pi/4}}{\sqrt{8\pi k_{\bot}|\boldsymbol{\rho}-\boldsymbol{\rho}_\mathrm{c}|}}
\oint\limits_C 
e^{jk_\bot
\hat s_\bot^\mathrm{c}\cdot (\boldsymbol{\rho}'-\boldsymbol{\rho}_\mathrm{c})}  \\
\bigg[
\boldsymbol{K}_\mathrm{e}(\boldsymbol{\rho}',z)
\times
\boldsymbol{k}_\mathrm{s}^\mathrm{c}  
-\omega\varepsilon
\boldsymbol{K}_\mathrm{m}(\boldsymbol{\rho}',z) 
+ 
\,\hat{n}'\cdot \boldsymbol{H}(\boldsymbol{\rho}',z) \, \boldsymbol{k}_\mathrm{s}^\mathrm{c}  
\bigg] \,dl'
\end{multline}
and
\begin{multline}\label{eq:F-11}
\boldsymbol{E}(\boldsymbol{\rho},z)  \approx 
\frac{e^{- jk_{\bot}|\boldsymbol{\rho}-\boldsymbol{\rho}_\mathrm{c}|+j\pi/4}}{\sqrt{8\pi k_{\bot}|\boldsymbol{\rho}-\boldsymbol{\rho}_\mathrm{c}|}}
\boldsymbol{k}_\mathrm{s}^\mathrm{c}\times\oint\limits_C 
e^{jk_\bot
\hat s_\bot^\mathrm{c}\cdot (\boldsymbol{\rho}'-\boldsymbol{\rho}_\mathrm{c})}   \\ 
\displaystyle \left[
\frac{1}{\omega\varepsilon}
\boldsymbol{k}_\mathrm{s}^\mathrm{c}\times\boldsymbol{K}_\mathrm{e}(\boldsymbol{\rho}',z)
+
\boldsymbol{K}_\mathrm{m}(\boldsymbol{\rho}',z)
\right]\,dl'  
\end{multline}
\begin{multline}\label{eq:F-12}
\boldsymbol{H}(\boldsymbol{\rho},z) \approx 
\frac{e^{- jk_{\bot}|\boldsymbol{\rho}-\boldsymbol{\rho}_\mathrm{c}|+j\pi/4}}{\sqrt{8\pi k_{\bot}|\boldsymbol{\rho}-\boldsymbol{\rho}_\mathrm{c}|}}
\boldsymbol{k}_\mathrm{s}^\mathrm{c}\times\oint\limits_C 
e^{jk_\bot
\hat s_\bot^\mathrm{c}\cdot (\boldsymbol{\rho}'-\boldsymbol{\rho}_\mathrm{c})}   \\ 
\displaystyle  \left[
\frac{1}{\omega\mu}
\boldsymbol{k}_\mathrm{s}^\mathrm{c}\times\boldsymbol{K}_\mathrm{m}(\boldsymbol{\rho}',z)
 -\boldsymbol{K}_\mathrm{e}(\boldsymbol{\rho}',z)
\right]\,dl'.
\end{multline}
It is seen that \eqref{eq:ST-7}--\eqref{eq:F-12} describe 
conical
waves with the phase center at $\boldsymbol{\rho}=\boldsymbol{\rho}_\mathrm{c}$. Furthermore, the fields $\boldsymbol{E}$ and $\boldsymbol{H}$ have no components in the direction of the vector $\boldsymbol{k}_\mathrm{s}^\mathrm{c}$, which is apparent from
the Schelkunoff-Franz line-equivalence relations \eqref{eq:F-11} and \eqref{eq:F-12}.

Finally, when the observation point is removed at a distance which is much greater than the distance between the phase center and the origin of the employed coordinate frame, i.e. $\rho\gg\rho_\mathrm{c}$, where $\rho=|\boldsymbol{\rho}|$ and $\rho_\mathrm{c}=|\boldsymbol{\rho}_\mathrm{c}|$, the following simplifications become valid:
\begin{equation}
\hat s_\bot^\mathrm{c} \approx \frac{\boldsymbol{\rho}}{\rho}=\hat\rho,~~~
|\boldsymbol{\rho}-\boldsymbol{\rho}_\mathrm{c}|\approx \rho-\hat\rho\cdot\boldsymbol{\rho}_\mathrm{c},
\end{equation}
implying that
\begin{equation}
\frac{e^{-jk_{\bot}|\boldsymbol{\rho}-\boldsymbol{\rho}_\mathrm{c}|+jk_\bot
\hat s_\bot^\mathrm{c}\cdot (\boldsymbol{\rho}'-\boldsymbol{\rho}_\mathrm{c})}}{\sqrt{|\boldsymbol{\rho}-\boldsymbol{\rho}_\mathrm{c}|}}\approx 
\frac{e^{-jk_{\bot}\rho + jk_\bot
\hat \rho\cdot \boldsymbol{\rho}'}}{\sqrt{\rho}},
\end{equation}
which makes \eqref{eq:ST-7}--\eqref{eq:F-12} to appear as 
conical
waves emanated from the 
$z$ axis
of the coordinate system.

The behavior of fields at such distances is fully described by 
far-field coefficients
\(F_{E}(\phi)\) and \(F_{H}(\phi)\), 
which are  defined by the relations (e.g. \cite[Sec. 3.4.4.]{OT}): 
\begin{eqnarray}
E_{z}\left( \boldsymbol{\rho},z \right) & \approx & 
\sqrt{\frac{2}{\pi k_{\bot}\rho}}F_{E}(\varphi)e^{- jk_{z}z - jk_{\bot}\rho - j3\pi/4}
 \label{eq:14a}   \\
H_{z}\left( \boldsymbol{\rho},z \right) & \approx & 
\sqrt{\frac{2}{\pi k_{\bot}\rho}}F_{H}(\varphi)e^{- jk_{z}z - jk_{\bot}\rho - j3\pi/4}.
 \label{eq:14b} 
\end{eqnarray}

Considering \eqref{eq:Ez} and \eqref{eq:Hz} in the limit $\rho\to+\infty$ and using 
\eqref{eq:exp} and 
the large-argument approximations of the Hankel functions in \eqref{eq:3}, \eqref{eq:8a} and \eqref{eq:8b} 
result in the formulas:
\begin{multline}\label{eq:16c} 
F_{E}(\varphi)  =  -\frac{k_\bot}{4}
\oint\limits_{C}
e^{jk_\bot\left( \hat{\rho}\cdot\boldsymbol{\rho}' \right)} \\  
\left[\left( \hat{\rho}\cdot {\hat{n}}'\right)E_z(\boldsymbol{\rho}',0)  +
\frac{j}{k_\bot}\frac{\partial}{\partial n'}E_z(\boldsymbol{\rho}',0)
\right]\,
dl'
\end{multline}
\begin{multline}\label{eq:16d}
F_{H}(\varphi) = -\frac{k_\bot}{4}
\oint\limits_{C}
e^{jk_\bot\left( \hat{\rho}\cdot\boldsymbol{\rho}' \right)} \\  
\left[
\left( \hat{\rho}\cdot {\hat{n}}'\right)H_z(\boldsymbol{\rho}',0)+
\frac{j}{k_\bot}\frac{\partial}{\partial n'}H_z(\boldsymbol{\rho}',0)
\right]\,
dl'.
\end{multline}

An alternative form follows from \eqref{eq:13a} and \eqref{eq:13b} 
as
\begin{multline}\label{eq:16a}
F_{E}(\varphi)  =   
-\frac{k_\bot}{4}
\oint\limits_{C}
e^{jk_{\bot}\left( \hat{\rho}\cdot\boldsymbol{\rho}' \right)}
\bigg[
\left( \hat{\rho}\cdot {\hat{n}}'\right) E_{z}\left( \boldsymbol{\rho}',0\right) \\ 
%\displaystyle - 
-\frac{k}{k_\bot}ZH_{t'}\left( \boldsymbol{\rho}',0\right)  
+ 
\frac{k_z}{k_\bot}E_{n'}\left( \boldsymbol{\rho}',0 \right)
 \bigg]\, dl'
\end{multline}
\begin{multline}\label{eq:16b}
F_{H}(\varphi) =   
- \frac{k_\bot}{4}
\oint\limits_{C}
e^{jk_{\bot}\left( \hat{\rho}\cdot\boldsymbol{\rho}' \right)}
\bigg[ 
\left(\hat{\rho}\cdot {\hat{n}}'\right)H_{z}\left( \boldsymbol{\rho}',0 \right) \\  
+ 
\frac{k}{k_\bot}YE_{t'}\left( \boldsymbol{\rho}',0\right) +
\frac{k_z}{k_\bot}H_{n'}\left( \boldsymbol{\rho}',0 \right)
 \bigg]\, dl',
\end{multline}
where $Z=\sqrt{\mu/\varepsilon}$ and $Y=\sqrt{\varepsilon/\mu}$ are the intrinsic impedance and admittance of the medium outside $C$.

Yet another form results from formulas \eqref{eq:F-5} and \eqref{eq:F-6} as
\begin{multline}\label{eq:16e}
F_{E}(\varphi)  =    
-\frac{k_\bot}{4}
\oint\limits_{C}
e^{jk_{\bot}\left( \hat{\rho}\cdot\boldsymbol{\rho}' \right)}
\bigg[ 
\left( \hat{\rho}\cdot{\hat{n}}'\right) E_{z}\left( \boldsymbol{\rho}',0\right) \\  
-
\frac{k_\bot}{k} ZH_{t'}\left( \boldsymbol{\rho}',0\right) 
 - \frac{k_z}{k}\left( \hat{\rho}\cdot {\hat{t}}'\right) 
ZH_z\left( \boldsymbol{\rho}',0 \right)
\bigg]\, dl'
\end{multline}
\begin{multline} \label{eq:16f}
F_{H}(\varphi) =  
- \frac{k_\bot}{4}
\oint\limits_{C}
e^{jk_{\bot}\left( \hat{\rho}\cdot\boldsymbol{\rho}' \right)}
\bigg[ 
\left( \hat{\rho}\cdot {\hat{n}}'\right) H_{z}\left( \boldsymbol{\rho}',0\right)   \\ 
+
\frac{k_\bot}{k} YE_{t'}\left( \boldsymbol{\rho}',0\right) 
 + \frac{k_z}{k}\left( \hat{\rho}\cdot {\hat{t}}'\right) 
YE_z\left( \boldsymbol{\rho}',0 \right)
\bigg]\, dl'.
\end{multline}

Equations \eqref{eq:16c}--\eqref{eq:16f} describe the far field of the sources enclosed by 
$\Gamma$.
The fields in the integrands are taken at $z=0$ since their dependence on $z$ is explicitly included in \eqref{eq:14a} and \eqref{eq:14b}.
Representations \eqref{eq:16c}--\eqref{eq:16f}
differ in the presence of the normal derivative, normal and tangential field components but 
despite their different appearance, describe identical 
far-field coefficients.

In the case of scattering problems, the scattering ability of cylindrical scatterers is described, depending on polarization, by the scattering widths $w_E$ and $w_H$. The incident field is assumed to be the plane waves 
\begin{eqnarray}
E_{z}^{\mathrm{inc}}  & = & E_{0z} e^{jk_\bot \rho \cos(\varphi-\varphi_0) - jk_{z}z}  
\label{17aa}  \\
H_{z}^{\mathrm{inc}}  & = & H_{0z}e^{jk_\bot \rho \cos(\varphi-\varphi_0) - jk_{z}z} 
\label{17bb} 
\end{eqnarray}
for the TM and TE polarization, respectively. 
The incidence direction is described by the polar angle $\varphi_0$ in the $x$-$y$ plane and
by the angle $\beta$ such that $k_z=k\cos\beta$, $k_\bot=k\sin\beta$ and $\beta=0$ corresponds to
the incidence in the direction of positive $z$ axis.
The far-field coefficients
are related to the scattering widths, 
and depending on the polarization of the incident and scattered waves, four polarization cases are distinguished,
\begin{equation}\label{eq:w1}
w_{EE}  =    \frac{4k}{k_\bot^2}\left| \frac{F_{E}}{E_{0z}} \right|^{2},~
w_{HE}  =    \frac{4k}{k_\bot^2}\left| \frac{F_{H}}{YE_{0z}} \right|^{2} 
\end{equation}
with $H_{0z}=0$ and 
\begin{equation}\label{eq:w2}
w_{EH}  =  \frac{4k}{k_\bot^2}\left| \frac{F_{E}}{ZH_{0z}} \right|^{2},~
w_{HH}  =  \frac{4k}{k_\bot^2}\left| \frac{F_{H}}{H_{0z}} \right|^{2}. 
\end{equation}
with $E_{0z}=0$.
The total scattering, extinction and absorption widths can also be expressed through 
the bistatic scattering widths $F_E$ and $F_H$ (e.g. \cite[Sec. 3.4.4]{OT}).

\section{Numerical checks and illustrations}\label{sec:numerical-illustrations}
The line-equivalence relations 
presented in 
Sections 
\ref{sec:ER-1}, \ref{sec:ER-2},
\ref{sec:Stratton-Chu} and \ref{sec:Franz}
as well as the formulas for the 
far-field coefficients
from Section 
\ref{sec:far-field}
are exact
analytical results. 
In order to verify their validity, a number of numerical checks 
have been carried out by using 
Wolfram Mathematica.
The following aspects have been considered:
\begin{enumerate}
\item radiation problems (primary sources inside $C$) and scattering problems (primary sources outside $C$); 
\item integration curve $C$ smooth (circle) and with corner points (square); 
\item fields $\boldsymbol{E}$ and $\boldsymbol{H}$ 
and 
far-field coefficients
$F_E$ and $F_H$; 
\item observation point inside and outside $C$.
\end{enumerate}
In all test cases the formulas for the equivalence relations have been confirmed within the standard machine precision.

\subsection{Radiation from a line
source}\label{subsec:radiation-from-a-line-source}

In the first test case, the electromagnetic field 
\begin{multline}\label{eq:17}
E_z\left( \boldsymbol{\rho},z \right) = E_{0z}e^{-jk_zz} \\
\cdot H_{0}^{(2)}\left[ k_\bot\sqrt{\left( x - x_{0} \right)^{2} + \left( y - y_{0} \right)^{2}} \right]    
\end{multline}
radiated by an electric line current at $x=x_0$ and $y=y_0$ 
with $E_{0z}=1 \, \text{V/m}$ and several values of $k_z$ has been used.
The other components of the fields can be obtained from \eqref{eq:9a}--\eqref{eq:9d}.
The rest parameters have been as follows:
\(x_{0} = 0 \,\text{m}\), \(y_{0} = 0.5 \,\text{m}\),  \(k = 1 \,\text{m}^{-1}\).  
Without loss of generality, the checks have been limited to the case of TM polarization ($H_{0z}=0$) since all presented relations are electromagnetically dual.

\begin{figure}
    \centering
    \includegraphics[width=0.8\linewidth]{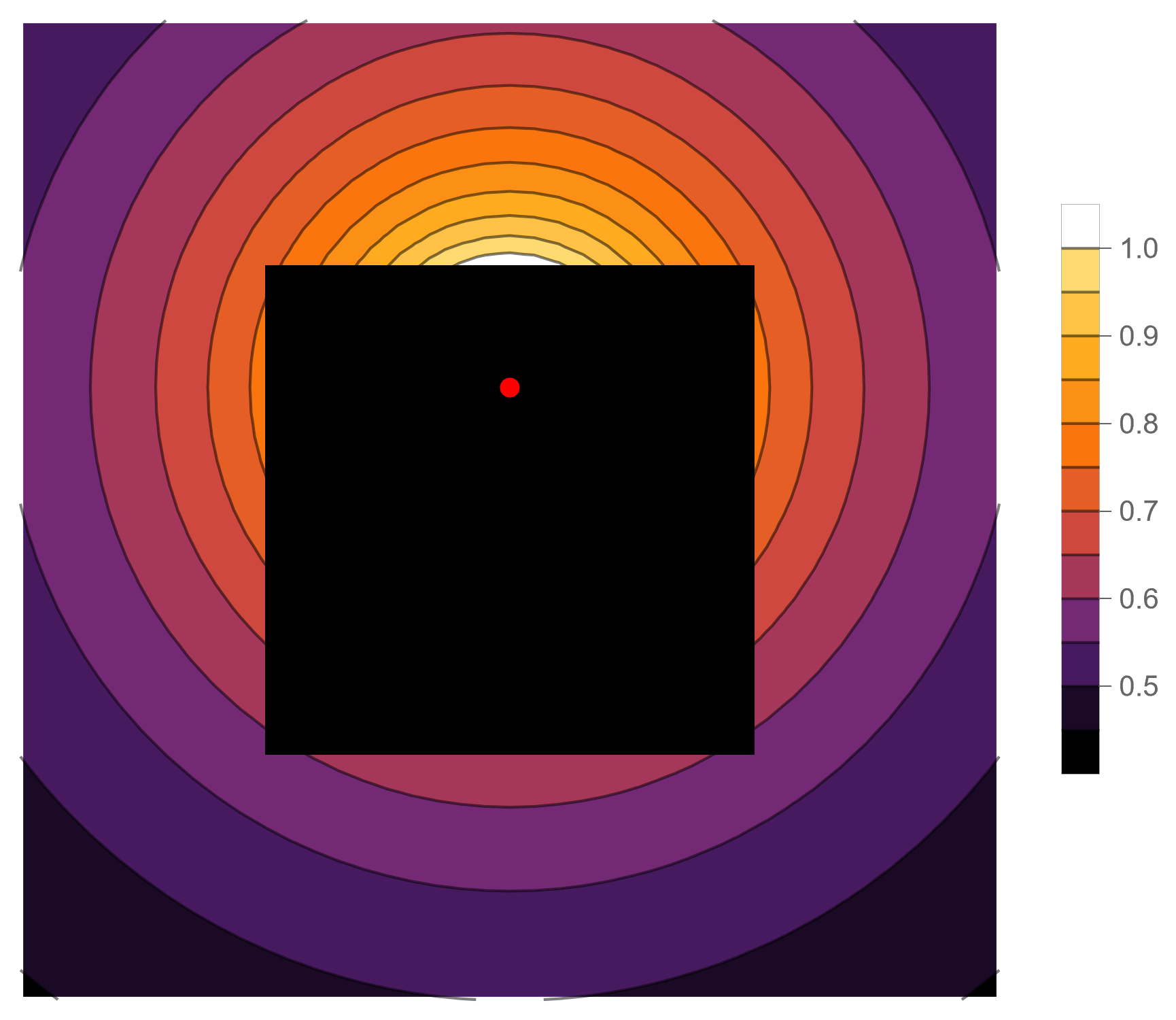}
    \caption{Amplitude of the real part of the field 
    \eqref{eq:17}
    due a line source 
    with $k_z=0$ and $x_0=0\, \text{m}$ and $y_0=0.5 \, \text{m}$
    (the position is indicated by the red
dot) computed by integrating Huygens' sources over a 
square enclosing contour.
The field is equal to zero inside the contour and
coincides with the field radiated by the source in the outside of the
contour.}
    \label{fig:2}
\end{figure}

Equations \eqref{eq:Ez} and \eqref{eq:Hz}, 
\eqref{eq:ST-3} and \eqref{eq:ST-4},
and \eqref{eq:F-7} and \eqref{eq:F-8}
have been tested. 
The enclosing line $C$ have been either a circle of radius $\rho'= 1 \,\text{m}$ centered at 
the origin of the coordinate system
or
a 
square
contour with
the corner points at \(\boldsymbol{\rho}_{1} = ( - 1\,\text{m}, - 1\,\text{m})\),
\(\boldsymbol{\rho}_{2} = (1\,\text{m}, - 1\,\text{m})\), \(\boldsymbol{\rho}_{3} = (1\,\text{m},1\,\text{m})\) and
\(\boldsymbol{\rho}_{4} = ( - 1\,\text{m},1\,\text{m})\). 
All formulations have 
recovered the 
fields $\boldsymbol{E}$ and $\boldsymbol{H}$
of the line source outside
the integration contour while giving zeros inside the integration contour.
Fig.~\ref{fig:2} shows the result of integration over the square contour $C$ for $|\mathrm{Re}\, E(x,y)|$ with $k_z=0$ in the region $-2\,\text{m}\le x,y \le 2 \,\text{m}$.
The results confirm the applicability of 
the presented line-equivalence formulas 
to radiation problems and
integration contours with and without corner points.

\subsection{Scattering from a dielectric cylinder}\label{subsec:scattering-from-cylinder}

Equivalence relations
can be applied to computing the fields scattered by cylindrical objects.
In this case, the sources are
secondary currents induced by an incident wave coming from outside of the enclosing 
surface $\Gamma$.
Consider, for example, the case of a dielectric circular cylinder illuminated by a plane
wave. The exact analytical solution is available (e.g. \cite[Sec. 6.5.1]{OT}, \cite[Sec. 4.1.3]{Ruck}) and can be used for checks of the equivalence formulas.
The fields and the scattering amplitudes can
be calculated either directly by using the exact series solution or
indirectly by using the equivalence relations. 

\begin{figure}
    \centering
    \includegraphics[width=0.75 \linewidth]{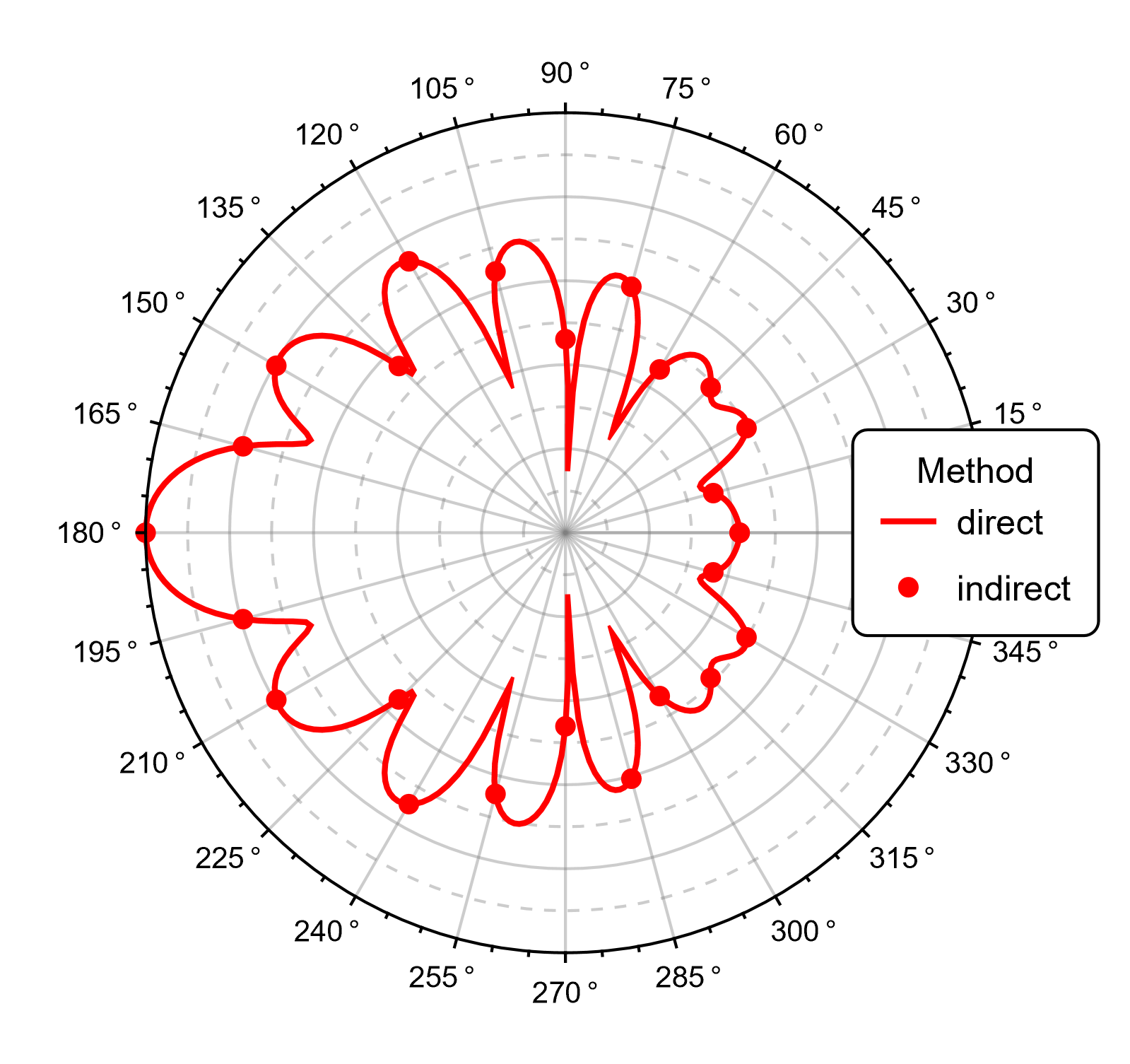}
        \caption{Log-polar plot of the bistatic scattering width $w_{EE}$ of a quartz glass circular cylinder with
 \(a = 0.05\,\text{m}\) for \(\beta = 45{^\circ}\) and
\(\varphi_{0} = 0{^\circ}\ \) at $80\,\text{GHz}$. 
The maximum value $w_{max}=5.34\, \text{m}$ is achieved on the outmost circle.
The results are presented on the dB scale
with the $5\,\text{dB}$ step between the grid circles.
}
    \label{fig:Wee}
\end{figure}

\begin{figure}
    \centering
    \includegraphics[width=0.75 \linewidth]{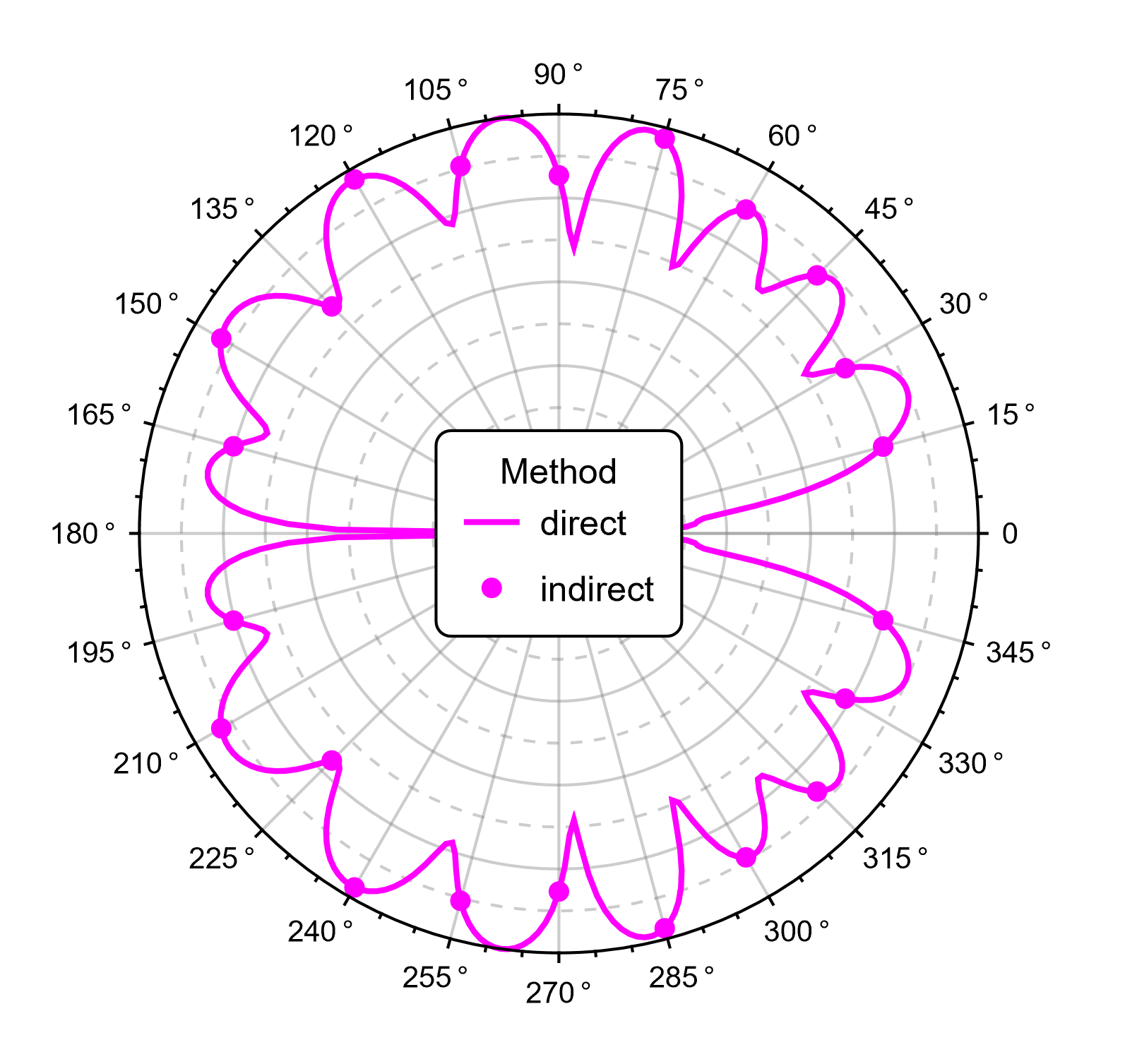}
    \caption{The same as in Fig.~\ref{fig:Wee} but for $w_{HE}$ ($w_{max}=0.11\,\text{m}$).}
    \label{fig:Whe}
\end{figure}

\begin{figure}
    \centering
    \includegraphics[width=0.75 \linewidth]{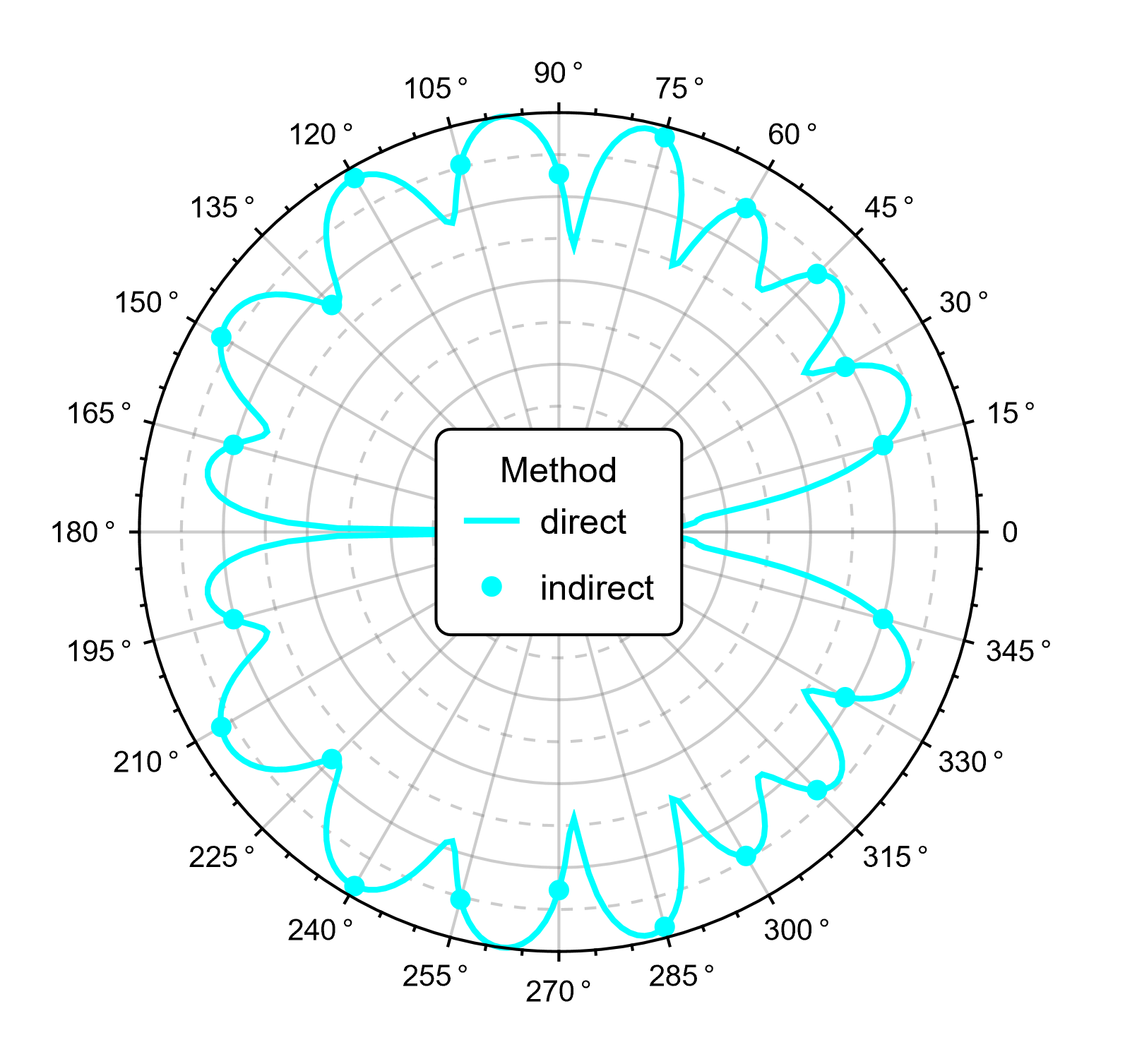}
    \caption{The same as in Fig.~\ref{fig:Wee} but for $w_{EH}$ ($w_{max}=0.11\,\text{m}$).
    The cross-pol plots coincide with each other, which is an intrinsic property of the solution for
    isotropic circular cylinders.}
    \label{fig:Weh}
\end{figure}

\begin{figure}
    \centering
    \includegraphics[width=0.75 \linewidth]{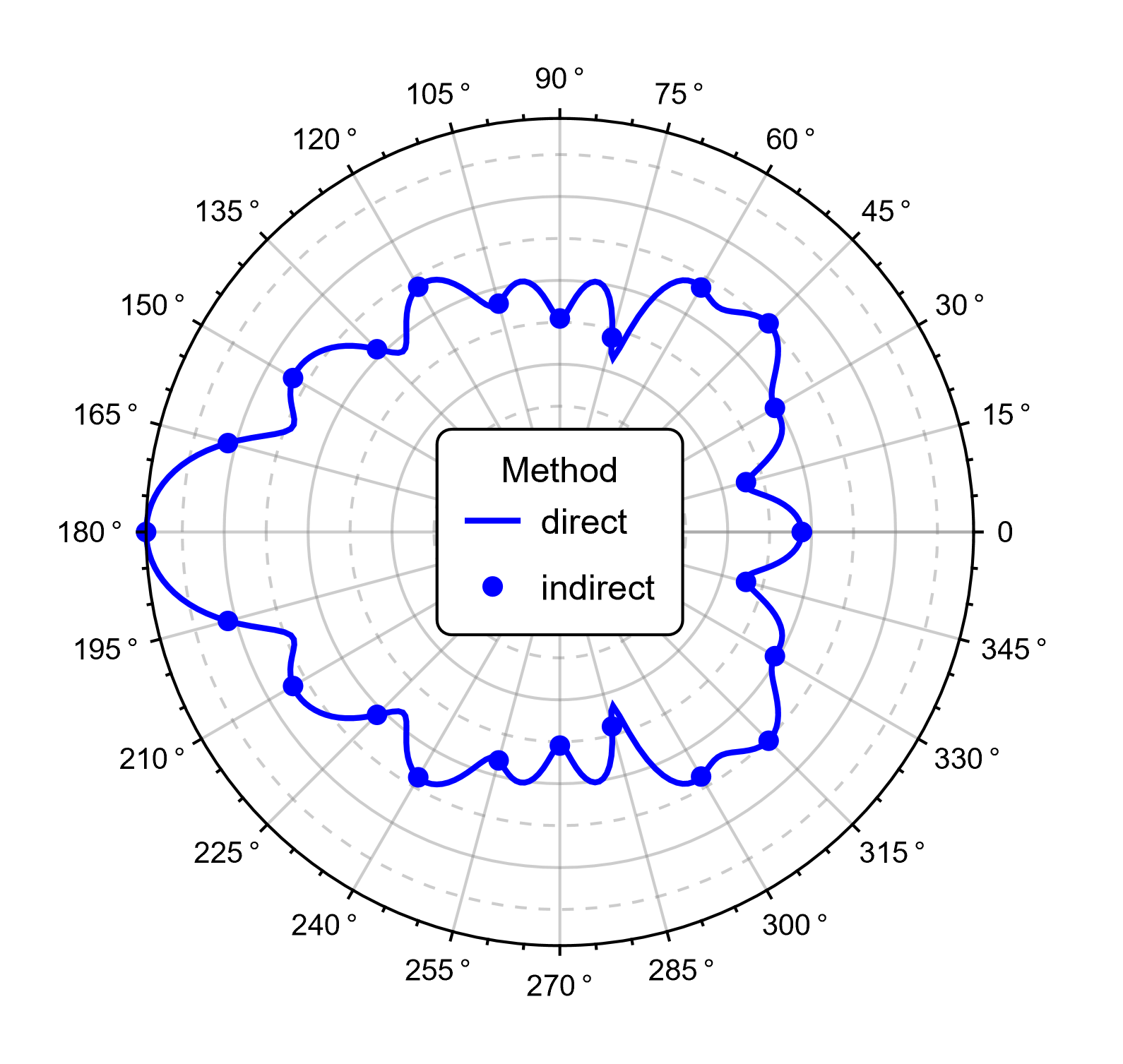}
    \caption{The same as in Fig.~\ref{fig:Wee} but  for $w_{HH}$ ($w_{max}=5.34\,\text{m}$).}
    \label{fig:Whh}
\end{figure}

Figures~\ref{fig:Wee} through \ref{fig:Whh} present the bistatic
scattering widths $w_{EE}$, $w_{HE}$, $w_{EH}$ and $w_{HH}$
as functions of the scattering
angle \(\varphi\) for a circular cylinder from quartz glass ($\epsilon_r'=3.81$, $\epsilon_r''= 0.003$, $\mu_r=1$)
with the radius \(a = 0.05\,\text{m}\) at $80\,\text{GHz}$.
The incidence direction is 
in the $x$-$z$ plane
(\(\varphi_{0} = 0{^\circ}\)) and makes the angle
\(\beta = 45{^\circ}\) with the cylinder axis.
The formulas \eqref{eq:16c}--\eqref{eq:16f} have been used.

The integration contour \(C\) has been chosen to be a
circle of the radius \(0.1\,\text{m}\)  with the center at the axis of the
circular cylinder. 
The suitable orthogonal
coordinate system is that of circular cylinder with
\(u_{1} = \rho\), \(u_{2} = \varphi\), \(h_{1} = 1\), 
\(h_{2} = \rho\) and $dl=\rho\,d\varphi$.
As expected, the direct and indirect methods have given identical
results, which is illustrated in Figs. \ref{fig:Wee} through \ref{fig:Whh}.
The agreement has also been observed for an observation point located at a finite distance
from the cylinder by using the field relations either
\eqref{eq:13a} and \eqref{eq:13b}, or \eqref{eq:F-5} and \eqref{eq:F-6}. 
Furthermore, the checks of the conventional
form \eqref{eq:Ez} and \eqref{eq:Hz} have also been conducted by inserting analytically evaluated normal derivatives.

\section{Conclusions}

Field equivalence relations for 
infinite cylindrical enclosing surfaces
and 
the $z$-dependence $\exp(-jk_zz)$ have been derived in the form of line integrals with the integration contour 
being the directrix of the cylindrical surface enclosing the source region.
The sources can be radiating and/or scattering 
bodies, 
arbitrarily shaped and
with arbitrary material constitution. 
Three versions of the field equivalence relation have been presented, analytically equivalent but differing in the presence of the specific field components (normal and tangential) or their normal derivative under the integration sign.

The conventional formulation (\eqref{eq:Ez}, \eqref{eq:Hz}, \eqref{eq:16c} and \eqref{eq:16d}) 
includes the normal derivative of the fields and can be inconvenient in numerical computations.
The formulations of the Stratton-Chu type 
(\eqref{eq:13a}, \eqref{eq:13b}, \eqref{eq:ST-3}, \eqref{eq:ST-4}, \eqref{eq:ST-5}, \eqref{eq:ST-6}, \eqref{eq:16c} and \eqref{eq:16d})
and
the Schelkunoff-Franz type (\eqref{eq:F-7}, \eqref{eq:F-8}, \eqref{eq:F-5}, \eqref{eq:F-6}, \eqref{eq:F-9}, \eqref{eq:F-10}, \eqref{eq:16e} and \eqref{eq:16f})
do not contain the normal derivative 
and can be advantageous when the fields are 
numerically
obtained.
Moreover, the relations of the Schelkunoff-Franz type
include only the tangential components and can be regarded as the rigorous electromagnetic formulation of Huygens's principle for infinite cylindrical 
enclosing surfaces.  

Furthermore, approximate versions of the line-equivalence relations in the intermediate and far zones have shown how Huygens' principle is to be understood for a cylindrical 
surface enclosing the sources: 
every element of the  
surface 
can be seen as a virtual linear source emanating a conical wave.

The relations have been specialized to the far-field limit ($\rho\to\infty$), and the corresponding expressions for the 
far-field coefficients
have been given, see \eqref{eq:16c}--\eqref{eq:16f}.

The derived expressions are applicable for a very general class of structures 
including those
that are infinite and periodic along one direction, while having a finite cross section in the transverse plane. The results allow   convenient calculations of fields at any point outside a cylindrical surface enclosing the structure by calculating a line integral of field values at the surface. For uniform structures, only one integration is sufficient. For periodic structures, calculations should be done for the fields of all significant Floquet harmonics. If only the far-field values are of interest and the period is smaller than half-wavelength, then also for periodic structures only the zero-order Floquet harmonic fields need to be integrated. 

For inhomogeneous and non-periodic structures, the line-equivalence relations are to be applied to all harmonics from the Fourier spectrum with respect to the $z$ coordinate.
If however
only the far field is of interest, then 
integration
over $k_z$ only from $-k$ to $k$, not from $-\infty$ to $\infty$,
is required 
as only the propagating part will contribute,
thus eliminating the
need to calculate fast-varying fields (large $k_z$).  
Note that in
the
conventional 3D surface integrals it is not easy to see how fine resolution for fields on the surface is enough. For periodic structures, far-field calculations need to be done only for a few (or even one) propagating Floquet harmonics of  the field expansion.

Importantly, the theory is also applicable when the propagation constant $k_z$ is complex-valued, allowing for studying leaky-wave regimes or active structures. In particular, the formulation entirely in terms of tangential components provides
a suitable framework for
describing reflection and transmission at  frequency-selective surfaces (FSS) and metasurface  structures (e.g., anomalous reflectors).  The intermediate-zone expressions \eqref{eq:ST-5}--\eqref{eq:F-10} are applicable, for example, in scenarios of millimeter-wave indoor communications, where the high-frequency approximation is valid, but the far-field conditions are usually not satisfied. The results can be also used  in numerical electromagnetics, offering fast means for near- to far-field transformations for solutions of 2D problems based on periodic boundary conditions.

The presented equivalence relations assume a bounded cross-section of the cylindrical geometry. They do not apply to cylindrical bodies with an unbounded cross-section, e.g. wedges.

\bibliographystyle{IEEEtran}
\bibliography{references}

\end{document}